\documentclass[a4paper,cite]{article}

\usepackage{latexsym,epsfig}

\usepackage[latin1]{inputenc}

\usepackage{graphicx}
\usepackage{color,pst-plot}

\usepackage{amsmath}
\usepackage {amsfonts,amssymb,amstext}
\usepackage{psfrag}
\usepackage{slashbox}

\newcommand{\beq}[1]{
\begin{equation}\label{#1}}
\newcommand{\eeq}{\end{equation}}
\newcommand{\bea}[1]{
\marginpar{\small\textsf{#1}}
\begin{eqnarray}\label{#1}}
\newcommand{\eea}{\end{eqnarray}}
\def\pv{\vec{p}_t}
\def\dv{\vec{\Delta}_t}

\def\ar{\alpha_\rho}
\def\mp{m_\pi}
\def\mr{m_\rho}

\def\bea{\begin{eqnarray}}
\def\beqa{\begin{eqnarray}}
\def\eea{\end{eqnarray}}
\def\eqa{\end{eqnarray}}
\def\beas{\begin{eqnarray*}}
\def\eeas{\end{eqnarray*}}
\def\beqas{\begin{eqnarray*}}
\def\eqas{\end{eqnarray*}}
\def\beq{\begin{equation}}
\def\eeq{\end{equation}}
\def\beqd{\begin{displaymath}}
\def\eeqd{\end{displaymath}}
\def\eqd{\end{displaymath}}

\def\beeq{\begin{eqnarray}} \def\eeeq{\end{eqnarray}}

\newcommand{\alb}{\bar{\alpha}}

\newcommand{\eq}{\end{equation}}

\newcommand{\be}{\begin{equation}}
\newcommand{\ee}{\end{equation}}

\DeclareMathAlphabet{\eusm}{U}{}{}{}
\SetMathAlphabet\eusm{normal}{U}{eus}{m}{n}
\SetMathAlphabet\eusm{bold}{U}{eus}{b}{n}
\DeclareMathAlphabet{\mathpzc}{OT1}{pzc}{m}{it}

\input epsf

\def\slashchar#1{\setbox0=\hbox{$#1$}
  \dimen0=\wd0
  \setbox1=\hbox{/} \dimen1=\wd1
  \ifdim\dimen0>\dimen1
     \rlap{\hbox to \dimen0{\hfil/\hfil}}
     #1
  \else
     \rlap{\hbox to \dimen1{\hfil$#1$\hfil}}
     /
  \fi}

\begin{document}

\begin{titlepage}

\begin{flushright}
CPHT-RR001.0110\\
LPT-10-07\\

\end{flushright}

\vspace*{0.2cm}
\begin{center}
{\Large {\bf Photoproduction of a  $\pi \rho_T$  pair  with a large invariant mass  and transversity generalized parton distribution
}}\\[2 cm]

{\bf M. El Beiyad}~$^{a, c}$, {\bf B. Pire}~$^a$, {\bf M. Segond}~$^b$, {\bf  L. Szymanowski}~$^{a,d}$ {\bf and S. Wallon}~$^{c,e}$\\[1cm]

$^a$  {\it Centre  de Physique Th{\'e}orique, \'Ecole Polytechnique, CNRS,
   91128 Palaiseau, France}\\[0.5cm]
$^b$ {\it  Institut f\"ur Theoretische
Physik, Universit\"at Leipzig,  D-04009 Leipzig, Germany}\\[0.5cm]
$^c$ {\it LPT, Universit\'e d'Orsay, CNRS, 91404 Orsay, France}\\[0.5cm]
$^d$ {\it Soltan Institute for Nuclear Studies, Warsaw, Poland}\\[0.5cm]
$^e$ {\it UPMC, Univ. Paris 06, Facult\'e de physique, 4 place Jussieu, 75252 Paris Cedex 05, France}
\end{center}

\vspace*{3.0cm}

\begin{abstract}
The chiral-odd transversity generalized parton distributions (GPDs) of the nucleon can be accessed experimentally through the exclusive photoproduction process $\gamma + N\ \to\  \pi+\rho +  N'$ , in the kinematics where the  meson pair has a large invariant mass and the final nucleon has a small transverse momentum, provided the vector meson is produced in a transversally polarized state. We  calculate perturbatively the scattering amplitude at leading order in $\alpha_s$. We build a simple model for the dominant transversity GPD $H_T(x,\xi ,t)$ based on the concept of double distribution. We estimate  the unpolarized differential cross section for this process  in the kinematics of the Jlab and COMPASS experiments. Counting rates show that the experiment
looks feasible with the real photon beam characteristics expected at 
JLab@12 GeV, and with the quasi
real photon beam in the COMPASS experiment.

  \end{abstract}
\vspace{1cm}

\end{titlepage}

\pagebreak
\thispagestyle{empty}
\mbox{}
\newpage
\setcounter{page}{1}
\section{ Introduction}
\label{Sec:Introduction}

Transversity quark distributions in the nucleon remain among the most unknown leading twist hadronic observables. This is mostly due to their chiral odd character which enforces their decoupling in most hard amplitudes. After the pioneering studies \cite{tra}, much work \cite{Barone} has been devoted to the exploration of many channels but experimental difficulties have challenged the most promising ones.

On the other hand, tremendous progress has been recently witnessed on the QCD description of hard exclusive processes, in terms of generalized parton distributions (GPDs) describing the 3-dimensional content of hadrons. Numerous experimental and theoretical reviews~\cite{review} exist now on this quickly developing subject. It is not an overstatement to stress that this activity is very likely to shed light on the confinement dynamics of QCD through the detailed understanding of the quark and gluon structure of hadrons.

Access to the chiral-odd transversity generalized parton distributions~\cite{defDiehl}, noted  $H_T$, $E_T$, $\tilde{H}_T$, $\tilde{E}_T$, has however turned out to be even more challenging~\cite{DGP} than the usual transversity distributions : one photon or one meson electroproduction leading twist amplitudes are insensitive to transversity GPDs. A possible way out is to consider higher twist contributions to these amplitudes \cite{liuti}, which however are beyond the factorization proofs and often plagued with end-point singularities. The strategy which we follow here, as initiated in Ref.~\cite{IPST,eps}, is to study the leading twist contribution to processes where more mesons are present in the final state; the hard scale which allows to probe the short distance structure of the nucleon is now the invariant mass of the meson pair, related to the large transverse momentum transmitted to  each final meson. In the example developed previously~\cite{IPST,eps}, the process under study was the high energy photo (or electro) diffractive production of two vector mesons, the hard probe being the virtual "Pomeron" exchange (and the hard scale being the virtuality of this pomeron), in analogy with the virtual photon exchange occuring in the deep inelastic electroproduction of a meson. A similar strategy has also been advocated recently in Ref.~\cite{kumano} to enlarge the number of processes which could be used to extract information on chiral-even GPDs.

The process we study here (\cite{DSPIN})
\begin{equation}
\gamma + N \rightarrow \pi^+ + \rho^0_T + N'\,,
\label{process1}
\end{equation}
is a priori sensitive to chiral-odd GPDs because of the chiral-odd character of the leading twist distribution amplitude of the transversally polarized $\rho$ meson. Its detailed study should not present major difficulties to modern detectors such as those developed for the 12 GeV upgrade of Jlab or for the Compass experiment at CERN. The estimated rate depends of course much on the magnitude of the chiral-odd generalized parton distributions. Not much is known about them, but model calculations have been developed in~\cite{eps} for the ERBL part and in Ref.~\cite{Sco,Pasq,othermodels}; moreover, a few moments have been computed on the lattice~\cite{lattice}. To supplement this and use the recent phenomenological knowledge acquired on the transversity quark distributions through single inclusive deep inelastic data, we propose in this paper a parametrization of the (dominant) transversity GPD $H_T^q$ based on the concept of double distributions.

Let us now explain how we factorize the amplitude of this process and what is the rational of this extension of the existing factorization proofs in the framework of QCD. The basis of our argument is two-folded.

\begin{figure}[h]
\begin{center}
\psfrag{z}{\begin{small} $z$ \end{small}}
\psfrag{zb}{\raisebox{0cm}{ \begin{small}$\bar{z}$\end{small}} }
\psfrag{gamma}{\raisebox{+.1cm}{ $\,\gamma$} }
\psfrag{pi}{$\,\pi$}
\psfrag{rho}{$\,\rho$}
\psfrag{TH}{\hspace{-0.2cm} $T_H$}
\psfrag{tp}{\raisebox{.5cm}{\begin{small}     $t'$       \end{small}}}
\psfrag{s}{\hspace{.6cm}\begin{small}$s$ \end{small}}
\psfrag{Phi}{ \hspace{-0.3cm} $\phi$}
\hspace{-0.7cm}
$\begin{array}{cc}
\hspace{.4cm}
\raisebox{.7cm}{\includegraphics[width=7cm]{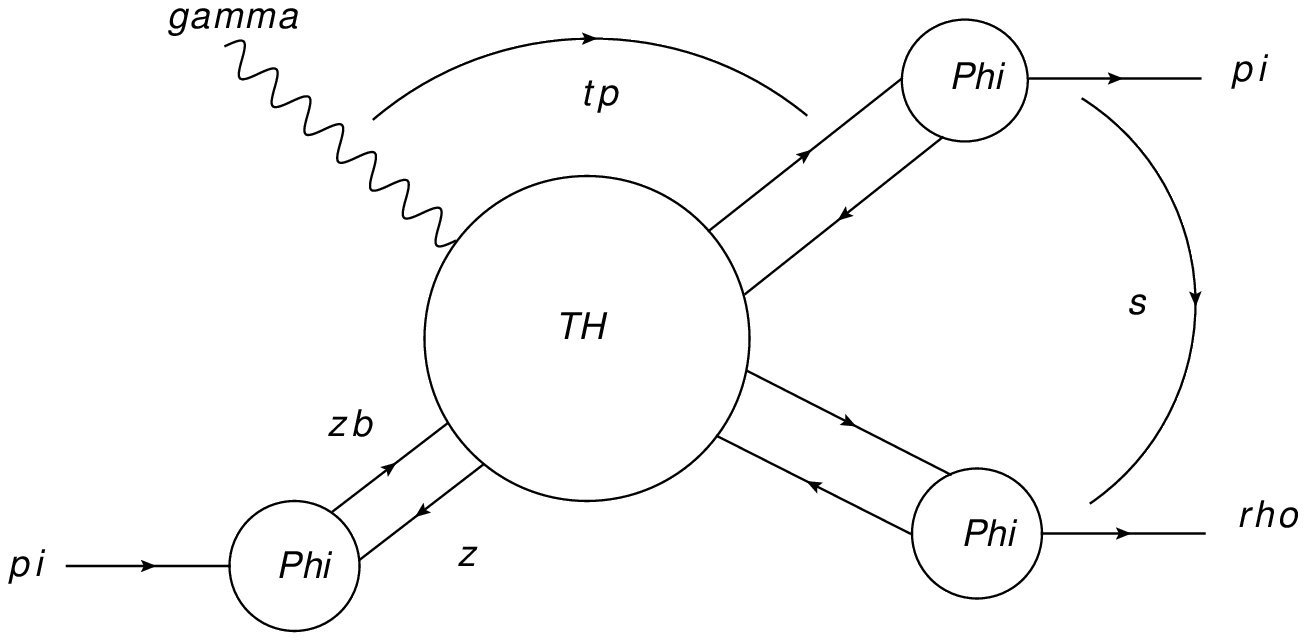}~~\hspace{0.6cm}} &
\psfrag{piplus}{$\,\pi^+$}
\psfrag{rhoT}{$\,\rho^0_T$}
\psfrag{M}{\hspace{-0.3cm} \begin{small} $M^2_{\pi \rho}$ \end{small}}
\psfrag{x1}{\hspace{-0.5cm} \begin{small}  $x+\xi $  \end{small}}
\psfrag{x2}{ \hspace{-0.2cm}\begin{small}  $x-\xi $ \end{small}}
\psfrag{N}{ \hspace{-0.4cm} $N$}
\psfrag{GPD}{ \hspace{-0.6cm}  $GPDs$}
\psfrag{Np}{$N'$}
\psfrag{t}{ \raisebox{-.1cm}{ \hspace{-0.5cm} \begin{small}  $t$  \end{small} }}
 \includegraphics[width=7cm]{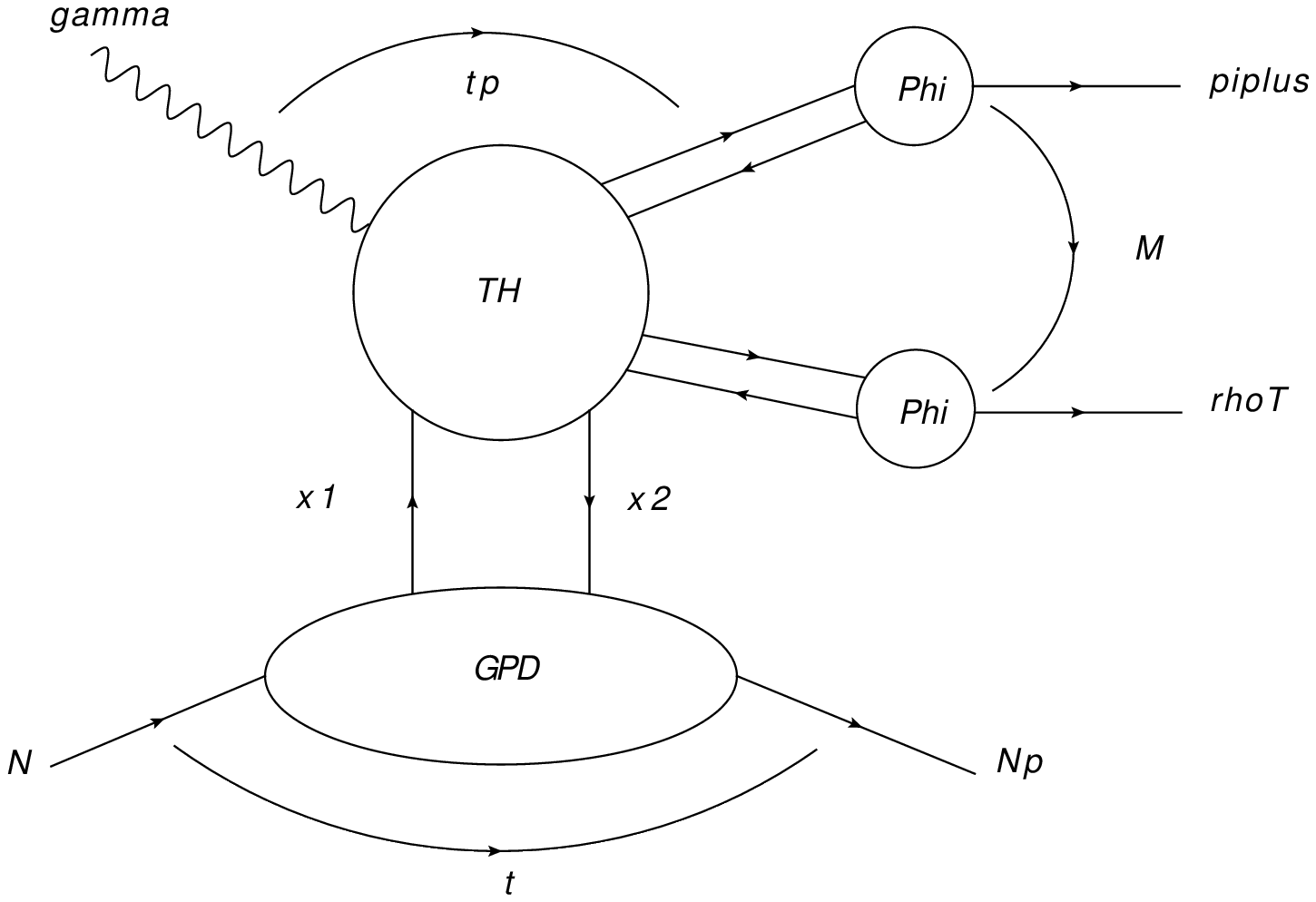} \\
 \\
 \hspace{-.5cm}(a) & \hspace{-1.5cm} (b)
 \end{array}
 $
\caption{a) Factorization of the amplitude for the process $\gamma + \pi \rightarrow \pi + \rho $ at large $s$ and fixed angle (i.e. fixed ratio $t'/s$); b) replacing one DA by a GPD leads to the factorization of the amplitude  for $\gamma + N \rightarrow \pi + \rho +N'$ at large $M_{\pi\rho}^2$\,.}
\label{feyndiag}
\end{center}
\end{figure}
\begin{itemize}
\item We use  the now classical proof of the factorization of exclusive scattering at fixed angle and large energy~ \cite{LB}. The amplitude for the process $\gamma + \pi \rightarrow \pi + \rho $ is written as the convolution of mesonic distribution amplitudes and a hard scattering subprocess amplitude $\gamma +( q + \bar q) \rightarrow (q + \bar q) + (q + \bar q) $ with the meson states replaced by collinear quark-antiquark pairs. This is described in Fig.~\ref{feyndiag}a. The demonstration of absence of any pinch singularities (which is the weak point of the proof for the generic case $A+B\to C+D$ ) has been proven in the case  of interest here~\cite{FSZ}.

\item We extract from the factorization procedure of the deeply virtual Compton scattering amplitude near the forward region the right to replace in Fig.~\ref{feyndiag}a the lower left meson distribution amplitude by a $N \to N'$ GPD, and thus get Fig.~\ref{feyndiag}b. Indeed the same collinear factorization property bases the validity of the leading twist approximation which either replaces the meson wave function by its distribution amplitude or the $N \to N'$ transition to its GPDs. A slight difference is that light cone fractions ($z, 1- z$) leaving the DA are positive, but the corresponding fractions ($x+\xi,\xi-x$) may be positive or negative in the case of the GPD. The calculation will show that this difference does not ruin the factorization property, at least at the order that we are working here.
\end{itemize}

One may adopt another point of view based on an analogy with the timelike Compton scattering
\begin{equation}
\gamma N  \to \gamma^* N' \to \mu^+ \mu^- N' \,,
\label{process2}
\end{equation}
where the lepton pair has a large squared  invariant mass $Q^2$, is instructive. This process has been thoroughly discussed~\cite{TCS} in the framework of the factorization of GPDs, and it has been proven that its amplitude was quite similar to the deeply virtual Compton scattering one, being dominated at lowest order by the handbag diagram amplitude convoluted with generalized quark distributions in the nucleon. There is no ambiguity  in this case for the definition of the hard scale, the photon virtuality $Q$ being the only scale present. Although the meson pair in the process  (\ref{process1}) has a more complex momentum flow, we feel justified to draw on this analogy to ascribe the role of the hard scale to the meson pair invariant squared mass. However, to describe the final state mesons by their distribution amplitudes (DAs), one needs in addition a large transverse momentum (and thus a large Mandelstam $t'$, see Fig.~\ref{feyndiag}b. Practically, we consider kinematics in which $|u'| \sim |t'| \sim |p_T^2| \sim M_{\pi \rho}^2 = (p_\pi +p_\rho)^2.$ We cannot prove, at the level of our study, that $M_{\pi \rho}^2$ is the most adequate hard scale. Indeed, applying a definite strategy to define a factorization scale requires at least a next to leading  (in the strong coupling) analysis \cite{BLM} and this is clearly a major work to be undertaken.

For both point of view, in order for the factorization of a partonic amplitude to be valid, and the leading twist calculation to be sufficient, one should avoid the dangerous kinematical regions where a small momentum transfer is exchanged in the upper blob, namely small $t' =(p_\pi -p_\gamma)^2$ or small $u'=(p_\rho-p_\gamma)^2$, and the regions where strong interactions between two hadrons in the final state are non-perturbative, namely where the invariant masses, $M^2_{\pi N'} = (p_\pi +p_{N'})^2$, $M^2_{\rho N'} = (p_\rho +p_{N'})^2$ and $M^2_{\pi\rho}$, are not large enough to suppress final state interactions. We will discuss the necessary minimal cuts to be applied to data before any attempt to extract the chiral odd GPDs.
However, although the ultimate proof of the validity of the factorization scheme proposed in this paper is based on comparison of the predictions with experimental data, on the theoretical side it requires to go beyond Born approximation considered here which is beyond the scope of the present work.

Our paper is organized as follows. In section \ref{Sec:Kinematics}, we clarify the  kinematics we are interested in and set our conventions. Then, in section \ref{Sec:Scattering_Amplitude}, we  describe the scattering amplitude of the process under study  in the framework of  QCD factorization. Section \ref{Sec:DD} is devoted to the presentation of our model chiral-odd GPDs. Section \ref{Sec:Cross Section and Rates} presents our results for  the unpolarized  differential cross section in the kinematics of two specific  experiments : quasi-real photon beams at JLab where $S_{\gamma N} \sim$ 14-20 GeV$^2$ and Compass at CERN where $S_{\gamma N} \sim$ 200 GeV$^2$.

As a final remark in this introduction, let us stress that our discussion applies as well to the case of electroproduction where a moderate virtuality of the initial photon may help to access the perturbative domain with a lower value of the hard scale $M_{\pi\rho}$.

\section{Kinematics}
\label{Sec:Kinematics}

We study the exclusive photoproduction of a transversely polarized vector meson and a pion on a  polarized or unpolarized proton target

\begin{equation}
\gamma(q) + N(p_1,\lambda) \rightarrow \pi(p_\pi) + \rho_T(p_\rho) + N'(p_2,\lambda')\,,
\label{process}
\end{equation}
 in the kinematical regime of large invariant mass $M_{\pi\rho}$ of the final meson pair and small momentum transfer $t =(p_1-p_2)^2$ between the initial and the final nucleons. Roughly speaking, these kinematics mean a moderate to large, and approximately opposite, transverse momentum of each  meson.
Our conventions  are the following. We decompose momenta on a Sudakov basis  as
\begin{equation}
\label{sudakov1}
k^\mu = a \, n^\mu + b \, p^\mu + k_\bot^\mu \,,
\end{equation}
with $p$ and $n$ the light-cone vectors
\begin{equation}
\label{sudakov2}
p^\mu = \frac{\sqrt{s}}{2}(1,0,0,1)\qquad n^\mu = \frac{\sqrt{s}}{2}(1,0,0,-1) \qquad p\cdot n = \frac{s}{2}\,,
\end{equation}
and
\begin{equation}
\label{sudakov3}
k_\bot^\mu = (0,k^x,k^y,0) \,, \qquad k_\bot^2 = -\vec{k}_t^2\,.
\end{equation}
The particle momenta read
\begin{equation}
\label{impini}
 p_1^\mu = (1+\xi)\,p^\mu + \frac{M^2}{s(1+\xi)}\,n^\mu~, \quad p_2^\mu = (1-\xi)\,p^\mu + \frac{M^2+\vec{\Delta}^2_t}{s(1-\xi)}n^\mu + \Delta^\mu_\bot\,, \quad q^\mu = n^\mu ~,
\end{equation}
\beqa
\label{impfinc}
p_\pi^\mu &=& \alpha \, n^\mu + \frac{(\vec{p}_t-\vec\Delta_t/2)^2+m^2_\pi}{\alpha s}\,p^\mu + p_\bot^\mu -\frac{\Delta^\mu_\bot}{2}~,\nonumber \\
 p_\rho^\mu &=& \alpha_\rho \, n^\mu + \frac{(\vec{p}_t+\vec\Delta_t/2)^2+m^2_\rho}{\alpha_\rho s}\,p^\mu - p_\bot^\mu-\frac{\Delta^\mu_\bot}{2}\,,
\eqa
with $\bar{\alpha} = 1 - \alpha$ and $M$, $m_\pi$, $m_\rho$ the masses of the nucleon, the pion and the $\rho$ meson.
From these kinematical relations it follows

\beq
\label{2xi}
2 \, \xi = \frac{(\pv -\frac{1}2 \dv)^2 + \mp^2}{s \, \alpha} +
\frac{(\pv +\frac{1}2 \dv)^2 + \mr^2}{s \, \ar}
\eq
and
\beq
\label{exp_alpha}
1-\alpha-\ar = \frac{2 \, \xi \, M^2}{s \, (1-\xi^2)} + \frac{\dv^2}{s \, (1-\xi)}\,.
\eq
The total center-of-mass energy squared of the $\gamma$-N system is
\begin{equation}
\label{energysquared}
S_{\gamma N} = (q + p_1)^2 = (1+\xi)s + M^2\,.
\end{equation}
$\xi$ is the skewedness parameter which can be written in terms of the $\tau$ variable used in lepton pair production,  as
\begin{equation}
\label{skewness}
\xi = \frac{\tau}{2-\tau} ~~~~,~~~~\tau = \frac{M^2_{\pi\rho}-t}{S_{\gamma N}-M^2}\,.
\end{equation}
On the nucleon side, the transferred squared momentum is
\begin{equation}
\label{transfmom}
t = (p_2 - p_1)^2 = -\frac{1+\xi}{1-\xi}\vec{\Delta}_t^2 -\frac{4\xi^2M^2}{1-\xi^2}\,.
\end{equation}
The other various Mandelstam invariants read
\begin{eqnarray}
\label{M_pi_rho}
s'&=& ~(p_\pi +p_\rho)^2 = ~M_{\pi\rho}^2= 2 \xi \, s \left(1 - \frac{ 2 \, \xi \, M^2}{s (1-\xi^2)}  \right) - \dv^2 \frac{1+\xi}{1-\xi}\,, \\
\label{t'}
- t'&=& -(p_\pi -q)^2 =~\frac{(\vec p_t-\vec\Delta_t/2)^2+\bar\alpha\, m_\pi^2}{\alpha} \;,\\
\label{u'}
- u'&=&- (p_\rho-q)^2= ~\frac{(\vec p_t+\vec\Delta_t/2)^2+(1-\alpha_\rho)\, m_\rho^2}{\alpha_\rho}
 \; ,
\end{eqnarray}
and
\beqa
\label{M_pi_N}
M_{\pi N'}^2 = s\left(1-\xi+ \frac{(\pv-\dv/2)^2+ m_\pi^2}{s\, \alpha}\right)
\left(\alpha + \frac{M^2 + \dv^2}{s \, (1-\xi)}  \right) - \left(\pv + \frac{1}2 \dv \right)^2\,,
\eqa
\beqa
\label{M_rho_N}
M_{\rho N'}^2 = s\left(1-\xi+ \frac{(\pv+\dv/2)^2+ \mr^2}{s\, \ar}\right)
\left(\ar + \frac{M^2 + \dv^2}{s \, (1-\xi)}  \right) - \left(\pv - \frac{1}2 \dv \right)^2\,.
\eqa

The hard scale $M^2_{\pi\rho}$ is the invariant squared mass of the ($\pi^+$, $\rho^0$) system.  The leading twist calculation of the hard part only involves the approximated kinematics in the generalized Bjorken limit: neglecting $\vec\Delta_\bot$ in front of $\vec p_\bot$ as well as hadronic masses, it amounts to
\beqa
\label{skewness2}
M^2_{\pi\rho} &\approx & \frac{\vec{p}_t^2}{\alpha\bar{\alpha}} \,,
\\
\ar &\approx& 1-\alpha \equiv \alb \,,\\
\tau &\approx &
\frac{M^2_{\pi\rho}}{S_{\gamma N}-M^2}\,,\\
-t' & \approx & \bar\alpha\, M_{\pi\rho}^2 \quad \mbox{ and } \quad -u'  \approx  \alpha\, M_{\pi\rho}^2 \,.
\eqa


The typical cuts that one should apply are $-t', -u' > \Lambda^2$ and
$M_{\pi N'}^2 = (p_\pi +p_{N'})^2 > M_R^2$, $M_{\rho N'}^2= (p_\rho +p_{N'})^2 > M_R^2$ where $\Lambda \gg \Lambda_{QCD}$
and $M_R$ is a typical baryonic resonance mass. This amounts to cuts in
$\alpha $ and $\bar\alpha$ at fixed $M_{\pi\rho}^2$, which can
be
 translated in terms of $u'$ at fixed $M_{\pi\rho}^2$ and $t$.
These conditions boil down to a safe kinematical domain $(-u')_{min} \leq -u' \leq (-u')_{max} $ which we will discuss in more details in Section \ref{Sec:Cross Section and Rates}.




In the following, we will choose as kinematical independent variables $t, u', M^2_{\pi \rho}\,.$

\section{The Scattering Amplitude}
\label{Sec:Scattering_Amplitude}

We now concentrate on the specific process
\begin{equation}
\gamma(q) + p(p_1,\lambda) \rightarrow \pi^+(p_\pi) + \rho^0_T(p_\rho) + n(p_2,\lambda')\,.
\label{process_pn}
\end{equation}
Let us start by recalling the non-perturbative quantities which enter the scattering amplitude of our process (\ref{process_pn}).
The transversity generalized parton distribution of a parton $q$ (here $q = u,\ d$) in the nucleon target at zero momentum transfer is defined by~\cite{defDiehl}
\beqa
&&<n(p_2,\lambda')|\, \bar{d}\left(-\frac{y}{2}\right)\sigma^{+j}\gamma^5 u \left(\frac{y}{2}\right)|p(p_1,\lambda)> \nonumber \\
&&= \bar{u}(p_2,\lambda')\,\sigma^{+j}\gamma^5u(p_1,\lambda)\int_{-1}^1dx\ e^{-\frac{i}{2}x(p_1^++p_2^+)y^-}H_T^{ud}(x,\xi,t)\,,
\label{defGPD}
\eqa
where $\lambda$ and $\lambda'$ are the light-cone helicities of the nucleons $p$ and $n$. Here $H^{ud}_T$ is the flavor non-diagonal  GPD
\cite{Mankiewicz:1997aa} which can be expressed in terms of diagonal
ones as
\beq
\label{defHud}
H^{ud}_T = H^{u}_T-H^{d}_T\,.
\eq

The chiral-odd light-cone DA for the transversely polarized meson vector $\rho^0_T$,  is defined, in leading twist 2, by the matrix element~\cite{defrho}

\begin{equation}
\langle 0|\bar{u}(0)\sigma^{\mu\nu}u(x)|\rho^0(p,\epsilon_\pm) \rangle = \frac{i}{\sqrt{2}}(\epsilon^\mu_{\pm}(p)\, p^\nu - \epsilon^\nu_{\pm}(p)\, p^\mu)f_\rho^\bot\int_0^1du\ e^{-iup\cdot x}\ \phi_\bot(u),
\label{defDArho}
\end{equation}
where $\epsilon^\mu_{\pm}(p_\rho)$ is the $\rho$-meson transverse polarization and with $f_\rho^\bot$ = 160 MeV.\\

The light-cone DA for the pion $\pi^+$ is defined, in leading twist 2, by the matrix element (see for example \cite{defpi})

\begin{equation}
\label{defDApion}
\langle 0|\bar{d}(z)\gamma^\mu\gamma^5u(-z)|\pi^+(p) \rangle = ip^\mu f_\pi\int_0^1du\ e^{-i(2u-1)p\cdot z}\ \phi_\pi(u),
\end{equation}
with $f_\pi$ = 131 MeV.
In our calculations, we use the asymptotic form of these DAs : $\phi_\pi(u) = \phi_\bot(u) = 6\,u\bar{u}$.

 We now pass to the computation of the scattering amplitude of the process (\ref{process_pn}).
 As the order of magnitude of the hard scale is greater than 1 GeV$^2$, it is possible to study it in the framework of QCD factorization, where the invariant squared mass of the ($\pi^+$, $\rho^0$) system $M^2_{\pi\rho}$ is taken as the factorization scale.

The amplitude gets contributions from each of the four twist 2 chiral-odd GPDs $E_T\,, H_T\,, \tilde{E}_T\,, \tilde{H}_T$. However, all of them but $H_T$ are accompanied by kinematical factors which vanish at $\vec{\Delta}_t=0\,.$
The contribution  proportional to $H_T$ is thus dominant in the small $t$ domain which we are interested in. We will thus restrict our study to this contribution, so that the whole $t-$dependence will come from the $t$-dependence of $H_T$, as we model in Sec.~\ref{Sec:DD}.
Note that within the collinear framework, the hard part is computed with $\vec{\Delta}_t=0$.

\begin{figure}[!h]
$\begin{array}{cc}
\psfrag{fpi}{$\,\phi_\pi$}
\psfrag{fro}{$\,\phi_\rho$}
\psfrag{z}{\begin{small} $z$ \end{small}}
\psfrag{zb}{\raisebox{-.1cm}{ \begin{small}$\hspace{-.3cm}-\bar{z}$\end{small}} }
\psfrag{v}{\begin{small} $v$ \end{small}}
\psfrag{vb}{\raisebox{-.1cm}{ \begin{small}$\hspace{-.3cm}-\bar{v}$\end{small}} }
\psfrag{gamma}{$\,\gamma$}
\psfrag{pi}{$\,\pi^+$}
\psfrag{rho}{$\,\rho^0_T$}
\psfrag{N}{$N$}
\psfrag{Np}{$\,N'$}
\psfrag{H}{\hspace{-0.2cm} $H^{ud}_T(x,\xi,t_{min})$}
\psfrag{hard}{\hspace{-0.2cm} $H^{ud}_T(x,\xi,t_{min})$}
\psfrag{p1}{\begin{small}     $p_1$       \end{small}}
\psfrag{p2}{\begin{small} $p_2$ \end{small}}
\psfrag{p1p}{\hspace{-0.4cm}  \begin{small}  $p_1'=(x+\xi) p$  \end{small}}
\psfrag{p2p}{\hspace{-0.2cm} \begin{small}  $p_2'=(x-\xi) p$ \end{small}}
\psfrag{q}{\begin{small}     $q$       \end{small}}
\psfrag{ppi}{\begin{small} $p_\pi$\end{small}}
\psfrag{prho}{\begin{small} $p_\rho$\end{small}}
\includegraphics[width=7.8cm]{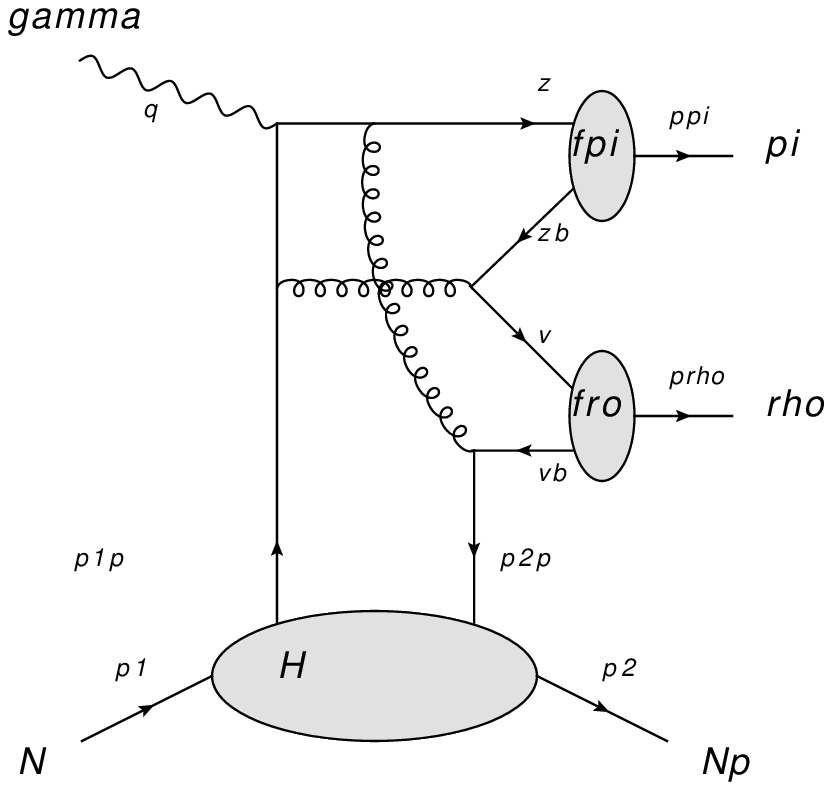}&\hspace{-0.5cm}
\psfrag{fpi}{$\,\phi_\pi$}
\psfrag{fro}{$\,\phi_\rho$}
\psfrag{z}{\begin{small} $z$ \end{small}}
\psfrag{zb}{\raisebox{-.1cm}{ \begin{small}$\hspace{-.3cm}-\bar{z}$\end{small}} }
\psfrag{v}{\raisebox{-.2cm}{\begin{small} $\hspace{-.2cm}-\bar{v}$ \end{small}}}
\psfrag{vb}{\raisebox{-.1cm}{ \begin{small}$v$\end{small}} }
\psfrag{gamma}{$\,\,\,\gamma$}
\psfrag{pi}{$\,\pi^+$}
\psfrag{rho}{$\,\rho^0_T$}
\psfrag{N}{$N$}
\psfrag{Np}{$\,N'$}
\psfrag{H}{\hspace{-0.2cm} $H^{ud}_T(x,\xi,t_{min})$}
\psfrag{p1}{\begin{small}     $p_1$       \end{small}}
\psfrag{p2}{\begin{small} $p_2$ \end{small}}
\psfrag{p1p}{\hspace{-0.4cm}  \begin{small}  $p_1'=(x+\xi) p$  \end{small}}
\psfrag{p2p}{\hspace{-0.2cm} \begin{small}  $p_2'=(x-\xi) p$ \end{small}}
\psfrag{q}{\begin{small}     $q$       \end{small}}
\psfrag{ppi}{\begin{small} $p_\pi$\end{small}}
\psfrag{prho}{\begin{small} $p_\rho$\end{small}}
\includegraphics[width=7.8cm]{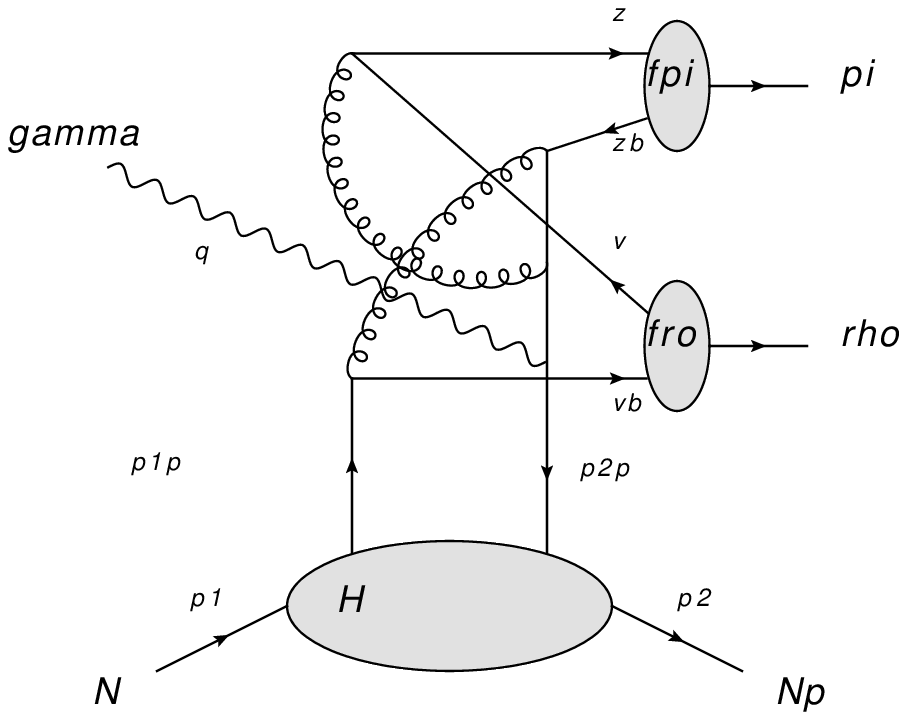}\\
\\ \hspace{-.5cm}(a) & \hspace{-.4cm}(b) \\
\end{array}$
\caption{Two representative diagrams with a photon $u$-quark coupling ($a$) and with a photon $d$-quark coupling ($b$).}
\label{feyndiageued}
\end{figure}

Thus we write the scattering amplitude of the process (\ref{process_pn}) in the factorized form :
\beq
\label{AmplitudeFactorized}
\mathcal{A}(t,M^2_{\pi\rho},p_T)  =\frac{1}{\sqrt{2}} \int_{-1}^1dx\int_0^1dv\int_0^1dz\ (T^u(x,v,z)-T^d(x,v,z)) \, H^{ud}_T(x,\xi,t)\Phi_\pi(z)\Phi_\bot(v)\,,
\eq
where $T^u$ and $T^d$ are the hard parts of the amplitude where the photon couples respectively to a $u$-quark (Fig.~\ref{feyndiageued}a) and to a $d$-quark (Fig.~\ref{feyndiageued}b). This decomposition, with the $\frac{1}{\sqrt{2}}$ prefactor, takes already into account that the  $\rho^0$-meson is described as $\frac{u\bar{u}-d\bar{d}}{\sqrt{2}}$.

\begin{figure}[!h]
\begin{center}
\psfrag{fpi}{$\,\phi_\pi$}
\psfrag{fro}{$\,\phi_\rho$}
\psfrag{z}{\begin{small} $z$ \end{small}}
\psfrag{zb}{\raisebox{-.1cm}{ \begin{small}$\hspace{-.3cm}-\bar{z}$\end{small}} }
\psfrag{v}{\begin{small} $v$ \end{small}}
\psfrag{vb}{\raisebox{-.1cm}{ \begin{small}$\hspace{-.3cm}-\bar{v}$\end{small}} }
\psfrag{gamma}{$\,\gamma$}
\psfrag{pi}{$\,\pi^+$}
\psfrag{rho}{$\,\rho^0_T$}
\psfrag{N}{$N$}
\psfrag{Np}{$\,N'$}
\psfrag{H}{\hspace{-0.2cm} $H^{ud}_T(x,\xi,t_{min})$}
\psfrag{p1}{\begin{small}     $p_1$       \end{small}}
\psfrag{p2}{\begin{small} $p_2$ \end{small}}
\psfrag{p1p}{\hspace{-0.1cm}  \begin{small}  $p_1'=(x+\xi) p$  \end{small}}
\psfrag{p2p}{\hspace{-0.2cm} \begin{small}  $p_2'=(x-\xi) p$ \end{small}}
\psfrag{q}{\begin{small}     $q$       \end{small}}
\psfrag{ppi}{\begin{small} $p_\pi$\end{small}}
\psfrag{prho}{\begin{small} $p_\rho$\end{small}}
\includegraphics[width=10cm]{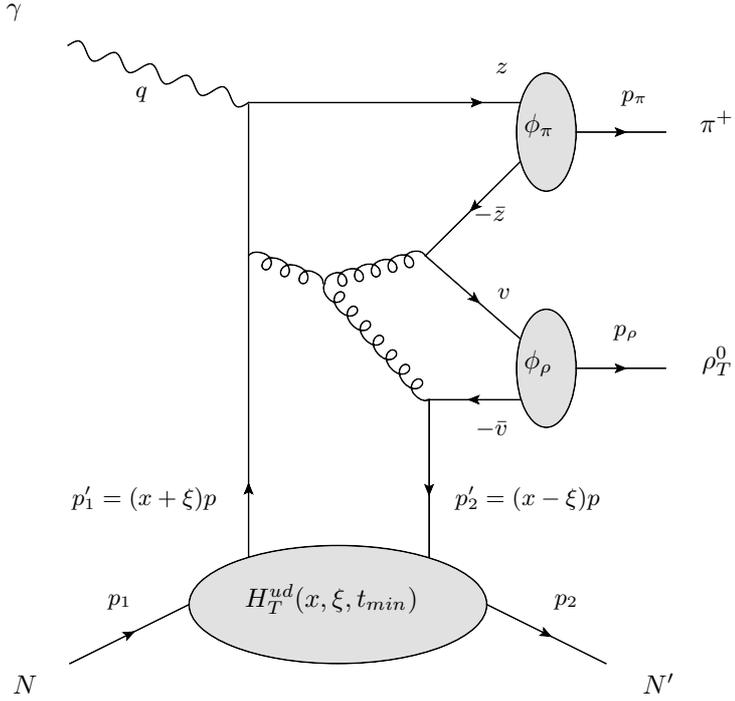}
\caption{Representative diagram with a 3 gluon vertex.}
\label{feyndiageu3g}
\end{center}
\end{figure}

For this process, one has two kinds of Feynman diagrams : some without (Fig.~\ref{feyndiageued}) and some with  a 3-gluon vertex (Fig.~\ref{feyndiageu3g}). 
In both cases, an interesting symmetry allows to deduce the contribution of some diagrams from other ones. This is examplified in Fig. \ref{feyndiageued}. The transformation rules
\begin{equation}
\label{symrules}
x\ \to\ -x \qquad u\ \to\ \bar{u} \qquad v\ \to\ \bar{v} \qquad e_{u}\ \to\ e_{d}
\end{equation}
relate the hard amplitude of Fig.~\ref{feyndiageued}b to the one of Fig.~\ref{feyndiageued}a. This reduces our task to the calculation of half the 62 diagrams involved in the process.\\
Let us sketch the main steps of the calculation on the specific example of the diagram of Fig.~\ref{feyndiageued}a,  in the Feynman gauge. Using the  notation $\slashchar{k} = k_\mu\gamma^\mu$, the amplitude reads :

\begin{eqnarray}
\label{hp2a_u}
T_{2a}^u(x,v,z) &=& Tr[(if_\pi\slashchar{p}_\pi\gamma^5)(-ig\gamma^\mu)\slashchar{F}(p'_2+\bar{v}p_\rho+zp_\pi)(ie_u\slashchar{\epsilon}(q))\slashchar{F}(p'_1-vp_\rho-\bar{z}p_\pi) \nonumber \\
&&(-ig\gamma^\nu)(\sigma^{\alpha\beta}\gamma^5)(-ig\gamma_\mu)(2i\sigma^{\sigma^*_\rho p_\rho}f_\rho^\bot)(-ig\gamma_\nu)] \nonumber \\
&\times&Tr_C[t^at^bt^at^b] \frac{1}{(8N_c)^2} \frac{1}{4N_c} G(p'_2+\bar{v}p_\rho)G(vp_\rho+\bar{z}p_\pi)\,,
\end{eqnarray}
where the fermion propagator is (we put all  quark masses to zero) :
\begin{equation}
\label{fermprop}
i\slashchar{F}(k) = \frac{i\slashchar{k}}{k^2+i\epsilon}\,,
\end{equation}
 and
\begin{equation}
\label{gluonprop}
-i g^{\mu \nu}G(k) = \frac{-i g^{\mu \nu}}{k^2+i\epsilon}
\end{equation}
is the gluonic propagator.
$Tr_C$ is the trace over color indices and the factors $\frac{1}{(8N_C)^2}$ and $\frac{1}{4N_C}$ come from Fierz decompositions. The corresponding expression for the diagram \ref{feyndiageued}b
\begin{eqnarray}
\label{hp2a_d}
T_{2b}^d(x,v,z) &=& Tr[(if_\pi\slashchar{p}_\pi\gamma^5)(-ig\gamma^\mu)(2i\sigma^{\sigma^*_\rho p_\rho}f_\rho^\bot)(-ig\gamma^\nu)(\sigma^{\alpha\beta}\gamma^5) \nonumber \\
&&(-ig\gamma_\mu)\slashchar{F}(p'_2+\bar{v}p_\rho+zp_\pi)(ie_d\slashchar{\epsilon}(q))\slashchar{F}(p'_1-vp_\rho-\bar{z}p_\pi)(-ig\gamma_\nu)] \nonumber \\
&\times&Tr_C[t^at^bt^at^b] \frac{1}{(8N_c)^2} \frac{1}{4N_c} G(-p'_1+vp_\rho)G(-\bar{v}p_\rho-zp_\pi)\nonumber \\
&=& \frac{iC_Fe_df^\bot_\rho f_\pi g^4\bar{z}}{32N_C^3s^2\bar{\alpha}[x-\xi-i\epsilon][x+\xi-i\epsilon]} \nonumber \\
&\times& \frac{\left[(\vec{N}_t\cdot\vec{\sigma}^*_{\rho t})(\vec{p}_t\cdot\vec{\epsilon}_{\gamma t})-(\vec{N}_t\cdot\vec{p}_t)(\vec{\epsilon}_{\gamma t}\cdot\vec{\sigma}*_{\rho t}) + \frac{2\alpha\xi-\bar{\alpha}}{2\alpha\xi+\bar{\alpha}}(\vec{N}_t\cdot\vec{\epsilon}_{\gamma t})(\vec{p}_t\cdot\vec{\sigma}^*_{\rho t})\right]}{zv\bar{v}[(\alpha\bar{z}+\bar{\alpha}v)(x+\xi-i\epsilon)-2\xi\bar{z}v]}
\end{eqnarray}
 justifies the symmetry we quoted a few lines above. Thus the hard part of the diagram \ref{feyndiageued}a is proportional to
\begin{equation}
\label{hp2a}
T_{2a}^u \propto \frac{1}{[(\slashchar{p}'_2+\bar{v}\slashchar{p}_\rho+z\slashchar{p}_\pi)^2+i\epsilon][(\slashchar{p}'_1-v\slashchar{p}_\rho-\bar{z}\slashchar{p}_\pi)^2+i\epsilon][(p'_2+\bar{v}p_\rho)^2+i\epsilon][(vp_\rho+\bar{z}p_\pi)^2+i\epsilon]}
\end{equation}
and  the  $i\epsilon$ prescription  in the 4 propagators  leads to the fact that the scattering amplitude gets both a real and an imaginary parts. Integrations over $v$ and $z$ have been done analytically whereas numerical methods are used for the integration over $x$. The first integration is rather straightforward. The second integration is more involved because of the presence of $i \epsilon$ terms inside the integrand, and in particular as an argument of logarithmic funtion, leading in the final result to appearance of imaginary parts. For example, the integration over $z$ of $T_{2b}^d$ requires to evaluate integrals of the type
\begin{equation}
\label{diffint}
\int_0^1 dz\ \frac{1}{2\,\xi\, z-\bar{\alpha}\,X}\log\left[\frac{\alpha Xz}{\bar{\alpha}\,X+z\,(\alpha X-2\xi)}\right]\,,
\end{equation}
where $X = x-\xi+i\epsilon$ contains all the dependence of the integrand on $i \epsilon$.\\
Nevertheless, since we have rewritten the $x$-dependence of propagators with the new variable X, it is possible to calculate this integral analytically without any problem. Thus the expression (\ref{diffint}) gives
\begin{equation}
\label{diffintresult}
\frac{\pi^2}{12\xi}+\frac{1}{2\xi}\mathrm{Li}_2\left[\left(1-\frac{2\xi}{\alpha X}\right)\left(1-\frac{2\xi}{\bar{\alpha} X}\right)\right]-\frac{1}{2\xi}\mathrm{Li}_2\left[1-\frac{2\xi}{\alpha X}\right]-\frac{1}{2\xi}\mathrm{Li}_2\left[1-\frac{2\xi}{\bar{\alpha} X}\right]\,.
\end{equation}

Lorentz invariance and the linearity of the amplitude with respect to the polarization vectors and with respect to the nucleons' spinors allow us to write the amplitude as :

\begin{eqnarray}
\mathcal{A} &=& (\epsilon^{*}_{\pm}(p_\rho)\cdot N^\bot_{\lambda_1\lambda_2})(\epsilon_{\gamma \bot}\cdot p_T)A' + (\epsilon^{*}_{\pm}(p_\rho)\cdot \epsilon_{\gamma \bot})(N^\bot_{\lambda_1\lambda_2}\cdot p_T)B' \nonumber\\
&+& (\epsilon^{*}_{\pm}(p_\rho)\cdot p_T)(N^\bot_{\lambda_1\lambda_2}\cdot \epsilon_{\gamma \bot})C' + (\epsilon^{*}_{\pm}(p_\rho)\cdot p_T)(N^\bot_{\lambda_1\lambda_2}\cdot p_T)(\epsilon_{\gamma \bot}\cdot p_T)D' \nonumber \\
&+& (\epsilon^{*}_{\pm}(p_\rho)\cdot p)(N^\bot_{\lambda_1\lambda_2}\cdot \epsilon_{\gamma \bot})E' + (\epsilon^{*}_{\pm}(p_\rho)\cdot p)(N^\bot_{\lambda_1\lambda_2}\cdot p_T)(\epsilon_{\gamma \bot}\cdot p_T)F',
\end{eqnarray}
where $A'$, $B'$, $C'$, $D'$, $E'$, $F'$ are scalar functions of $s$, $\xi$, $\alpha$ and $M^2_{\pi\rho}$, and the transverse polarization of $\rho$-meson 
\begin{equation}
\label{polarho}
\epsilon^{\mu}_{\pm}(p_\rho) = \left(\frac{\vec{p}_\rho\cdot\vec{e}_\pm}{m_\rho}\ ,\ \vec{e}_\pm + \frac{\vec{p}_\rho\cdot\vec{e}_\pm}{m_\rho(E_\rho + m_\rho)}\vec{p}_\rho \right)
\end{equation}
is expressed in terms of $\vec{e}_{\pm}=-\frac{1}{\sqrt{2}}(\pm 1,i,0)$. $\epsilon_{\gamma\bot}^\mu$ is the transverse polarization of the on-shell photon and
\begin{equation}
\label{polnucleon}
N^{\bot\mu}_{\lambda_1\lambda_2} = \frac{2i}{p\cdot n}g_\bot^{\mu\nu}\bar{u}(p_2,\lambda_2)\slashchar{n}\gamma_\nu\gamma^5u(p_1,\lambda_1)
\end{equation}
is the spinor dependent part which expresses the nucleon helicity flip with $g_\bot^{\mu\nu} = diag(0,-1,-1,0)$.\\
To be more precise, the expressions of this $2-$dimensional transverse vector read
  \begin{eqnarray}
  \label{polprocess}
N_{+\hat{x},+\hat{x}}^{\bot\mu} &=& -4i\sqrt{1-\xi^2}(0,1,0,0) \qquad N_{-\hat{x},+\hat{x}}^{\bot\mu} = 4\sqrt{1-\xi^2}(0,0,1,0) \\
N_{+\hat{x},-\hat{x}}^{\bot\mu} &=& -4\sqrt{1-\xi^2}(0,0,1,0) \qquad N_{-\hat{x},-\hat{x}}^{\bot\mu} = 4i\sqrt{1-\xi^2}(0,1,0,0)\,,
\end{eqnarray}
assuming that these nucleons are polarized along the $\hat{x}$ axis.\\
Since the DA of $\rho^0_T$ (Eq. (\ref{defDArho})) introduces the factor $\epsilon^\mu_{\pm}(p_\rho)\, p_\rho^\nu - \epsilon^\nu_{\pm}(p_\rho)\, p_\rho^\mu$, any term proportional to $p_\rho^\mu$ in its polarisation does not contribute to the amplitude. On may then replace
\begin{eqnarray}
\label{polarho2}
\epsilon^{\mu}_{\pm}(p_\rho) &\Rightarrow&
2 \bar \alpha\frac{ \vec p_t \cdot \vec e_\pm }{\bar \alpha^2 s +
\vec p_t^{\,2}} \left( p^\mu + n^\mu  \right) + (0,\vec e_\pm) \nonumber \\
&\Rightarrow& 2\bar{\alpha}\frac{\vec{p}_t\cdot\vec{e}_\pm}{\bar{\alpha}^2s + \vec{p}_t^{\,2}}\left[1 - \frac{\vec{p}_t^{\,2}}{\bar{\alpha}^2s}\right]p^\mu + 2\frac{\vec{p}_t\cdot\vec{e}_\pm}{\bar{\alpha}^2s + \vec{p}_t^{\,2}}
p_T^\mu + (0,\vec{e}_\pm).
\end{eqnarray}
Consequently, the amplitude of this process can be simplified as
\begin{eqnarray}
\label{ampl}
\mathcal{A} &=& (\vec{N}_t\cdot \vec{e}^{\,*}_\pm)(\vec{p}_t\cdot \vec{\epsilon}_{\gamma t}) A + (\vec{N}_t\cdot\vec{\epsilon}_{\gamma t})(\vec{p}_t\cdot\vec{e}^{\,*}_\pm) B \nonumber \\
&+& (\vec{N}_t\cdot\vec{p}_t) (\vec{\epsilon}_{\gamma t}\cdot\vec{e}^{\,*}_\pm) C + (\vec{N}_t\cdot\vec{p}_t) (\vec{p}_t\cdot\vec{\epsilon}_{\gamma t}) (\vec{p}_t\cdot \vec{e}^{\,*}_\pm) D
\end{eqnarray}
where $A$, $B$, $C$, $D$ are also scalar functions of $s$, $\xi$, $\alpha$ and $M^2_{\pi\rho}$.\\
The final result for each particular diagram is rather lenghty, and because of that we do not present
explicit final results for scalar functions  $A,$ $B$, $C$, $D$ of (\ref{ampl}).

\section{Transversity GPD and Double Distribution}
\label{Sec:DD}

\begin{figure}[h!]
$\begin{array}{cc}
\hspace{-.15cm}
\includegraphics[width=8.5cm]{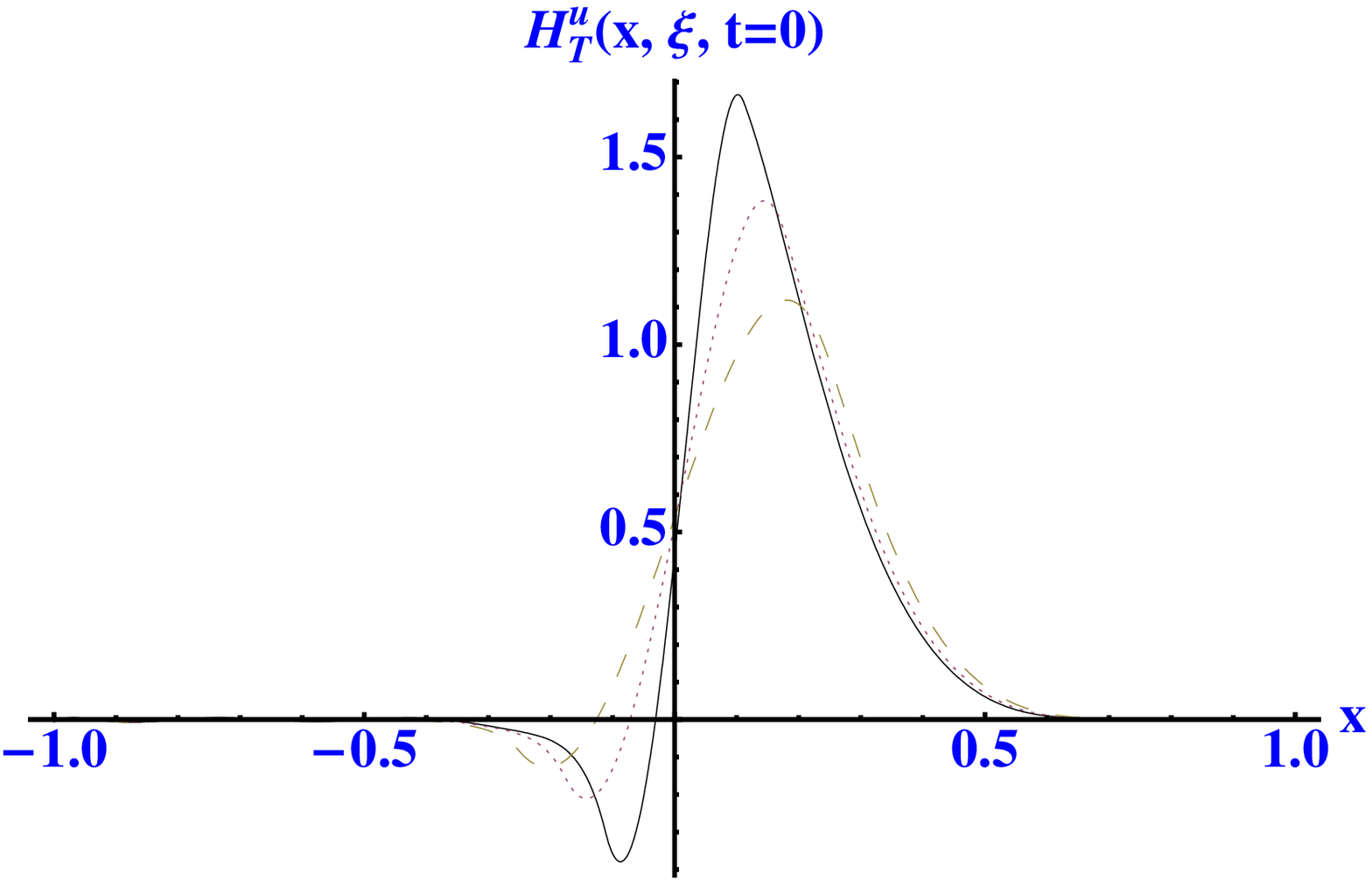}
&
\hspace{-1.9cm}
\includegraphics[width=8.5cm]{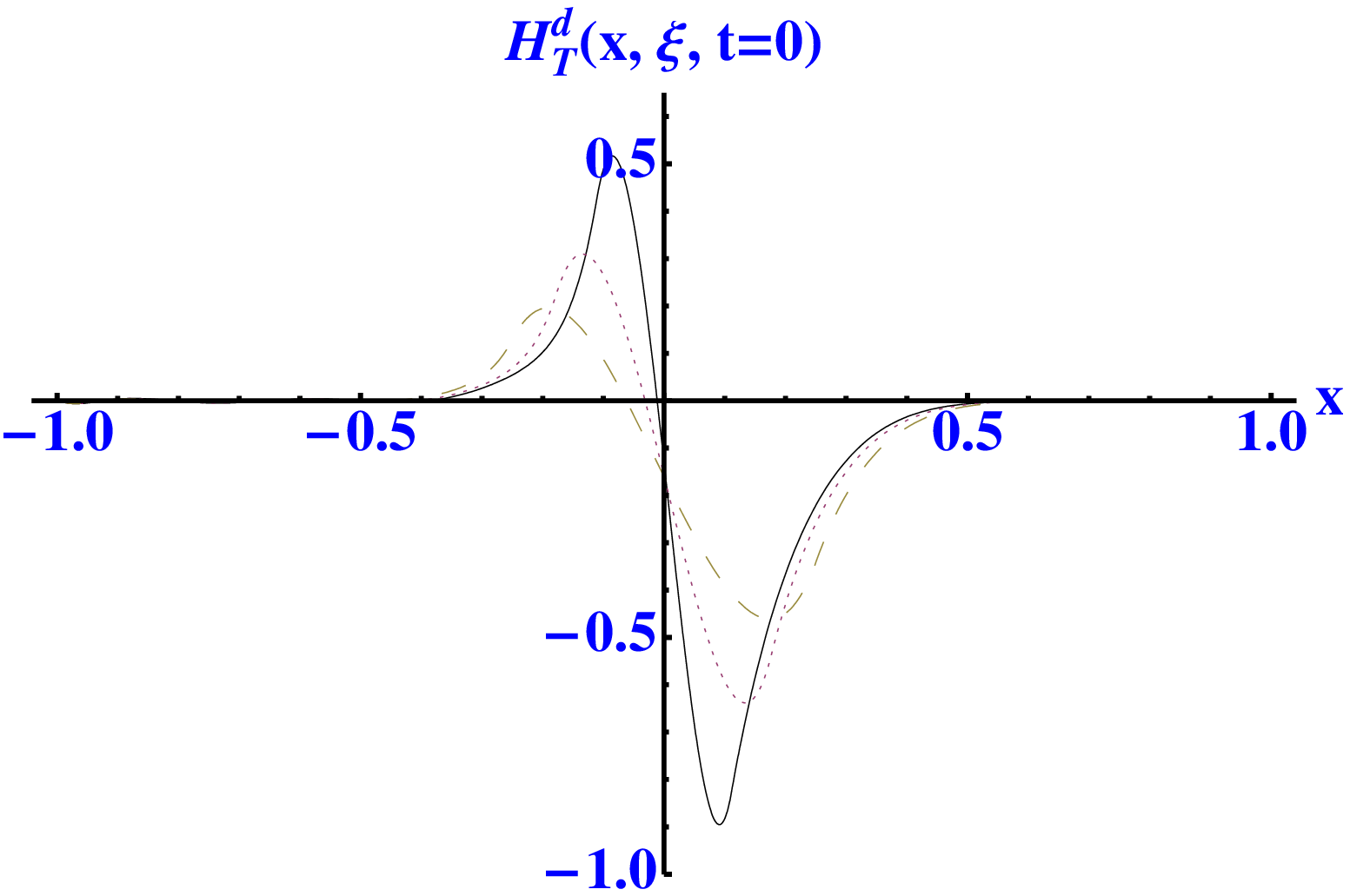}
\\
\\
  (a)    &  \hspace{-1.5cm}(b) \\
\end{array}$
\caption{Transversity GPD  $H_T^u(x,\xi,t=0)$ (a) and $H_T^d(x,\xi,t=0)$ (b) of the nucleon for $\xi$ = .111 (solid line), $\xi$ = .176 (dotted line), $\xi$ = .25 (dashed line), corresponding respectively to $M_{\pi \rho}^2/S_{\gamma N}$ equal to 4/20, 6/20 and 8/20.}
\label{HTu_d_JLab}
\end{figure}

\begin{figure}[h!]
$\begin{array}{cc}
\hspace{-.15cm}
\includegraphics[width=7.5cm]{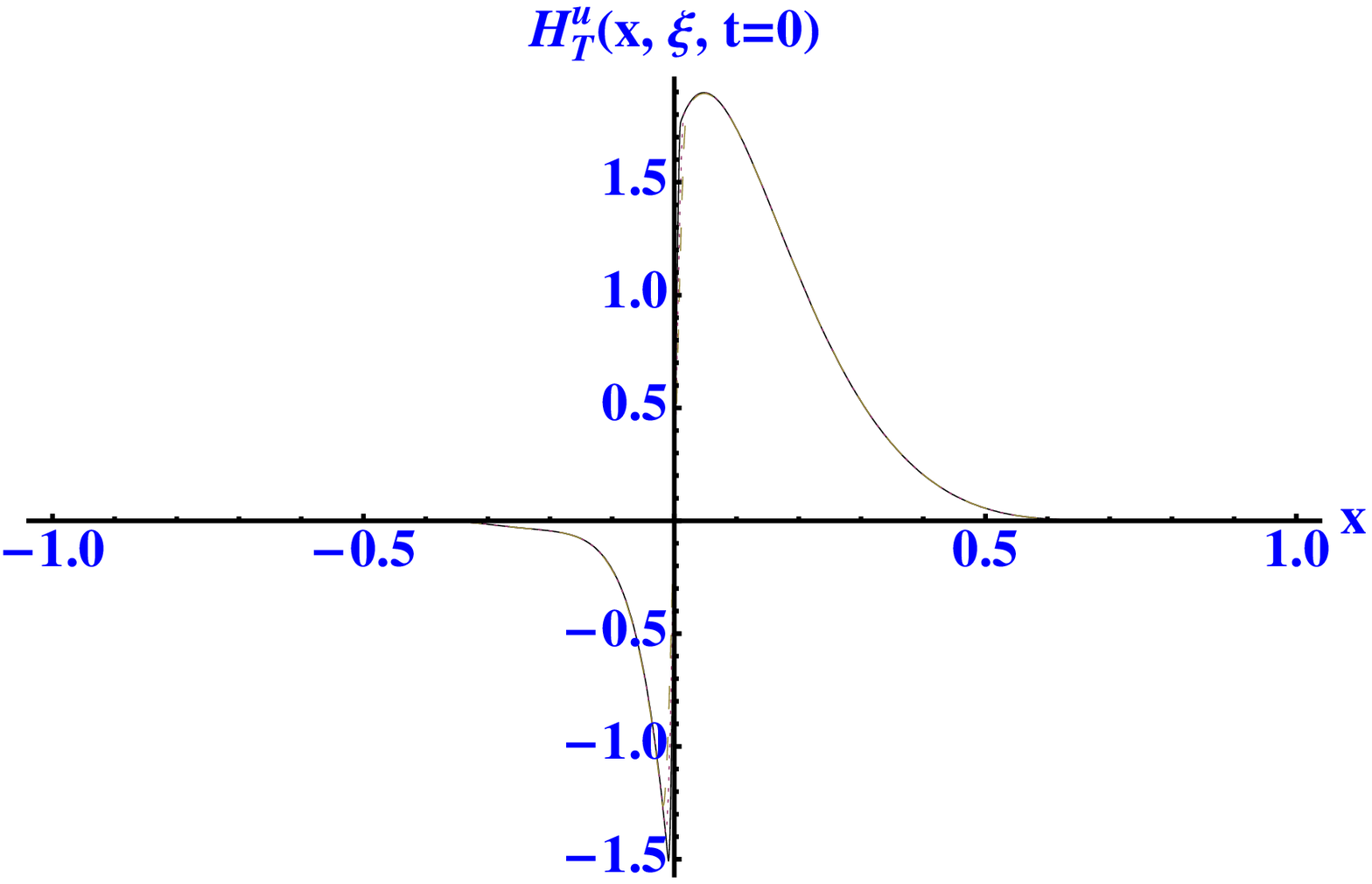}
&
\hspace{0cm}
\raisebox{-.22cm}{\includegraphics[width=7.5cm]{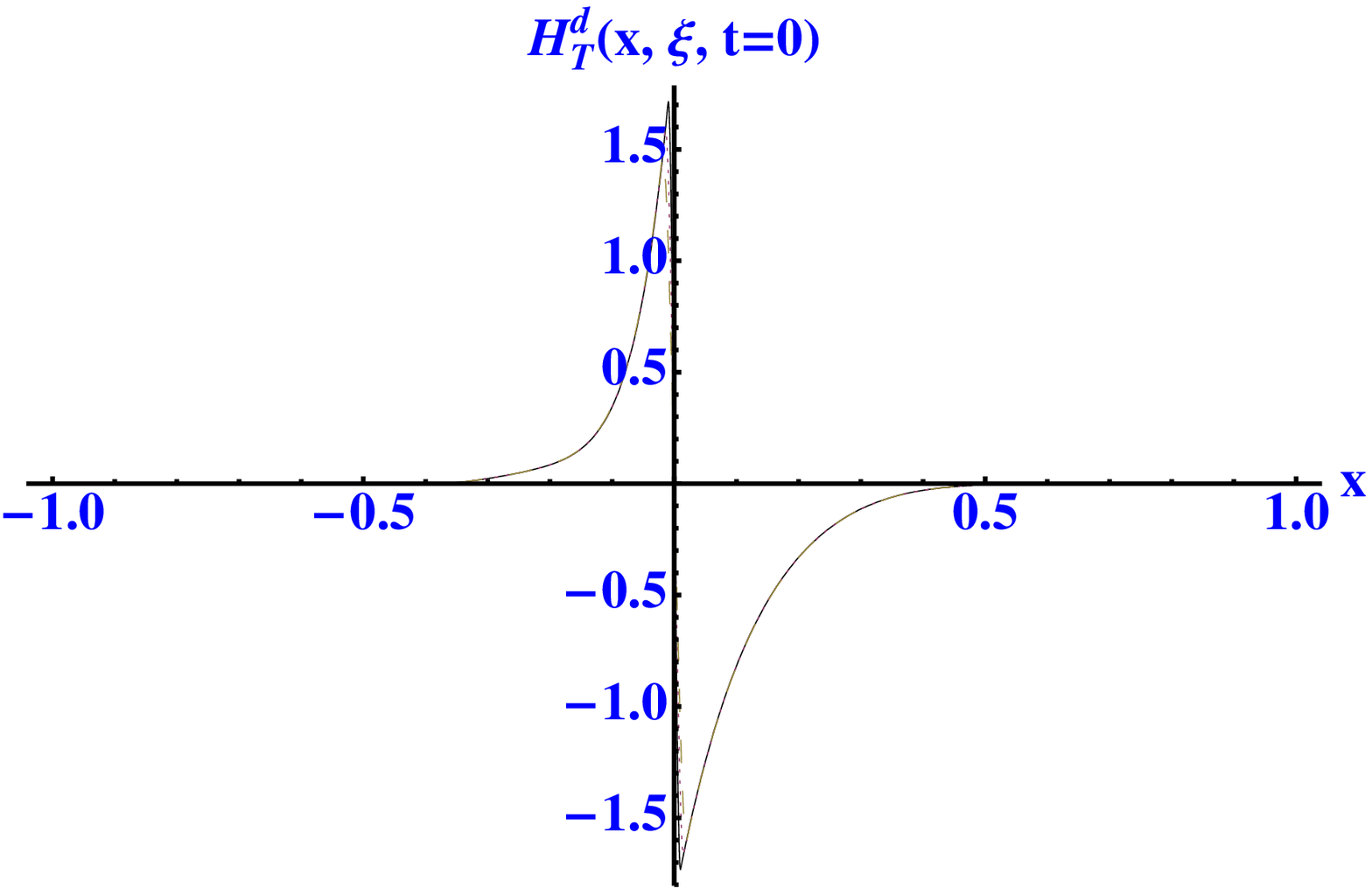}}
\\
\\
\hspace{-.15cm}
\includegraphics[width=7.5cm]{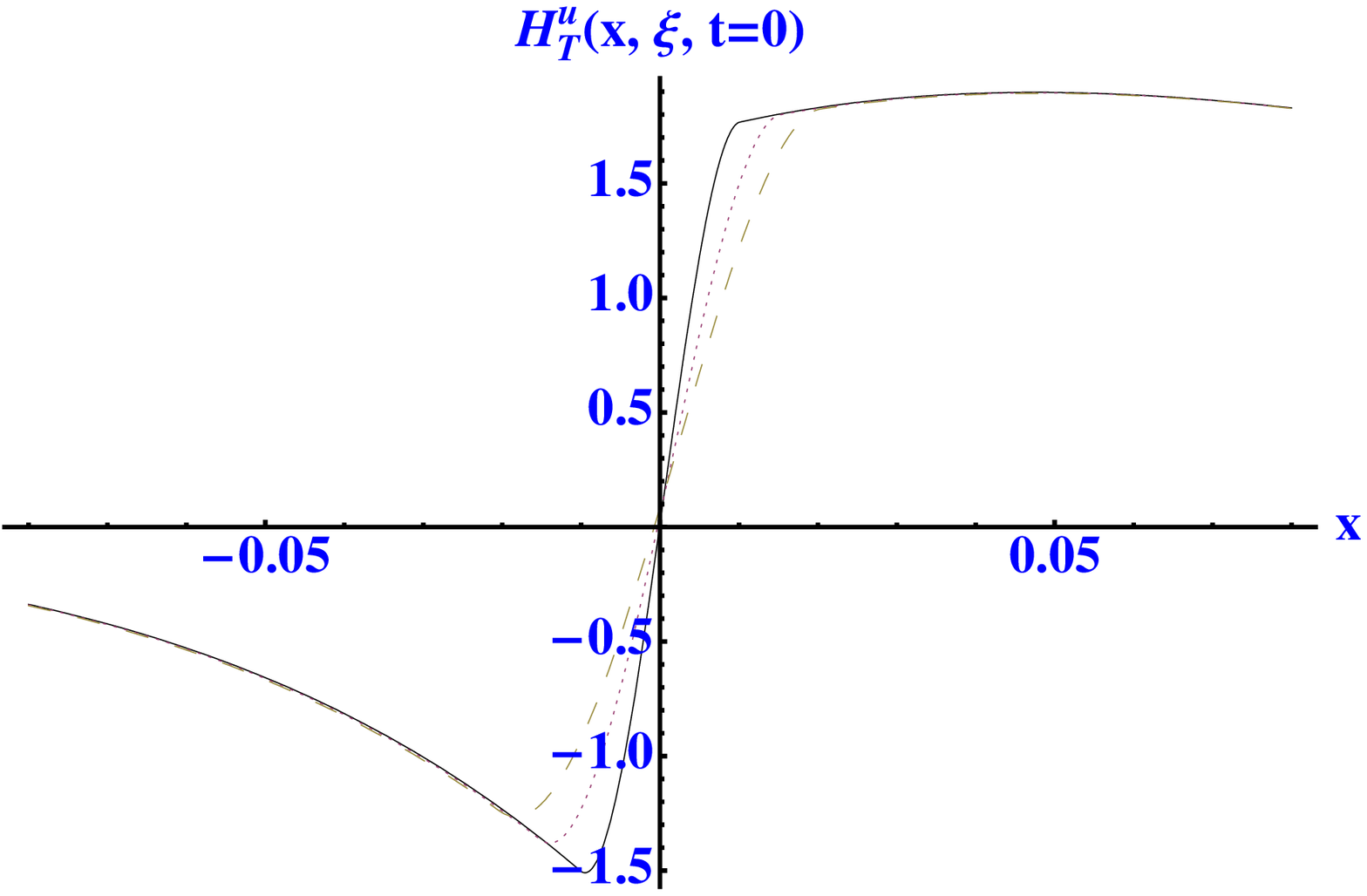}
&
\hspace{0cm}
\raisebox{-.23cm}{\includegraphics[width=7.5cm]{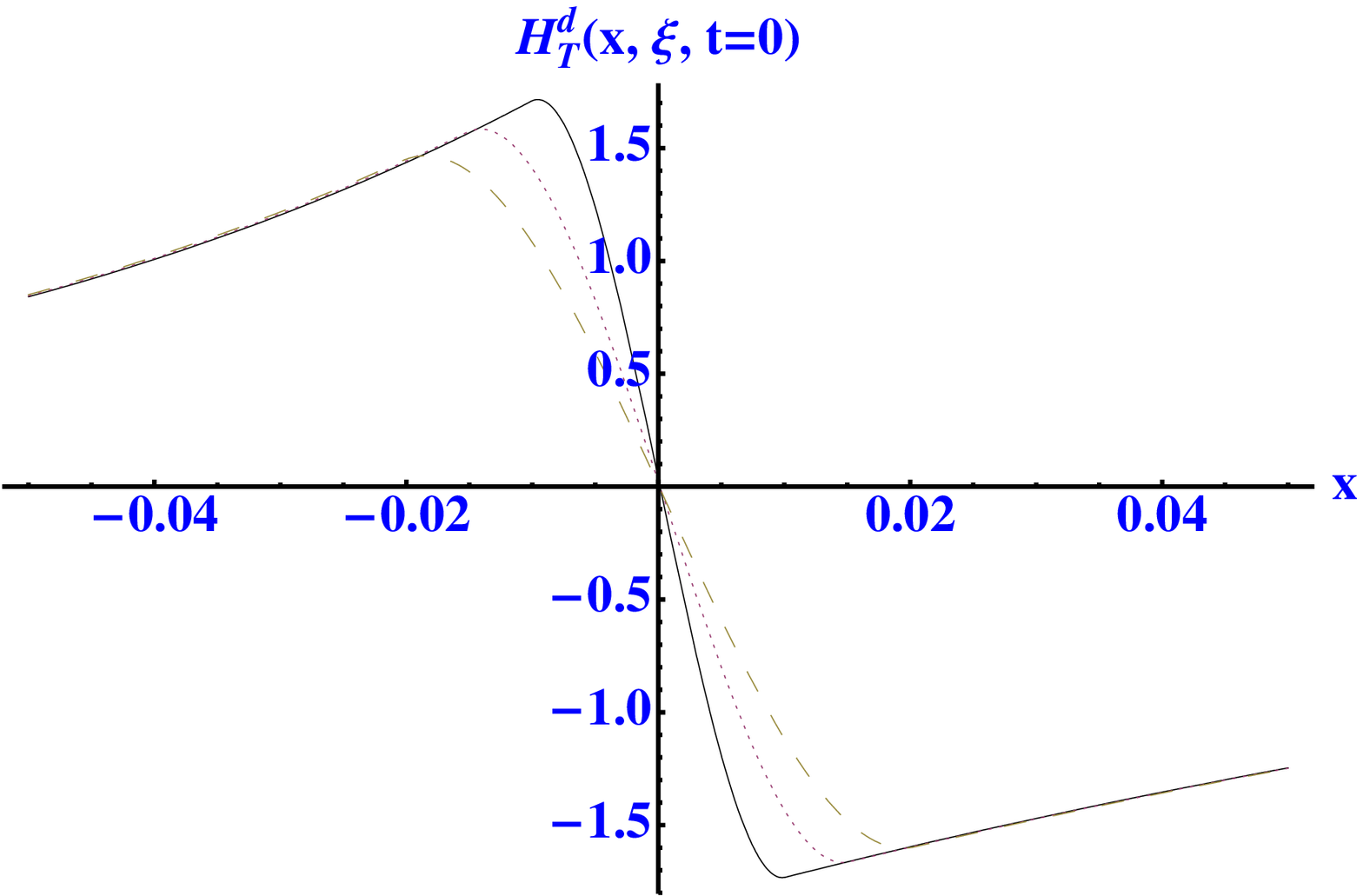}}
\\
\\
  (a)    &  (b) \\
\end{array}$
\caption{Transversity GPD  $H_T^u(x,\xi,t=0)$ (a) and $H_T^d(x,\xi,t=0)$ (b) of the nucleon for $\xi$ = .01 (solid line), $\xi$ = .015 (dotted line), $\xi$ = .02 (dashed line), corresponding respectively to $M_{\pi \rho}^2/S_{\gamma N}$ equal to 4/200, 6/200 and 8/200, the lower plots being blow-ups of the central part of the upper ones.}
\label{HTu_d_Compass}
\end{figure}

In order to get an estimate of the differential cross section of this process, we need to propose a model for the transversity GPD $H_T^q(x,\xi,t)$ ($q=u,\ d$). Contrary to what Enberg \textit{et al.} have done \cite{eps}, here we must get a parametrization in both ERBL ($]-\xi ; \xi[$) and DGLAP ($[-1 ; -\xi]\ \bigcup\ [\xi ; 1]$) $x-$domains.\\
We use the standard description of GPDs in terms of double distributions \cite{Rad}
\begin{equation}
\label{DDdef}
H_T^q(x,\xi,t=0) = \int_\Omega d\beta\, d\alpha\ \delta(\beta+\xi\alpha-x)f_T^q(\beta,\alpha,t=0) \,,
\end{equation}
where $f_T^q$ is the quark transversity double distribution and $\Omega = \{|\beta|+|\alpha| \leqslant 1\}$ is its support domain. Moreover we may add a D-term contribution, which is necessary to be completely  general while fulfilling the polynomiality constraints. Since adding a D-term is quite arbitrary and unconstrained, we do not include it in our parametrization. We thus propose a simple model for these GPDs, by writting
 $f_T^q$ in the form
\begin{equation}
\label{DD}
f_T^q(\beta,\alpha,t=0) = \Pi(\beta,\alpha)\,\delta \, q(\beta)\Theta(\beta) - \Pi(-\beta,\alpha)\,\delta \bar{q}(-\beta)\,\Theta(-\beta)\,,
\end{equation}
where $ \Pi(\beta,\alpha) = \frac{3}{4}\frac{(1-\beta)^2-\alpha^2}{(1-\beta)^3}$ is a profile function and $\delta q$, $\delta \bar{q}$ are the quark and antiquark transversity parton distribution functions (PDF).
The transversity GPD $H_T^q$ thus reads
\begin{eqnarray}
H_T^q(x,\xi,t=0) &=& \Theta(x>\xi)\int_{\frac{-1+x}{1+\xi}}^{\frac{1-x}{1-\xi}}dy\ \frac{3}{4}\frac{(1-x+\xi y)^2-y^2}{(1-x+\xi y)^3}\delta q(x-\xi y) \nonumber \\
&+&  \Theta(\xi>x>-\xi)\left[\int_{\frac{-1+x}{1+\xi}}^{\frac{x}{\xi}}dy\ \frac{3}{4}\frac{(1-x+\xi y)^2-y^2}{(1-x+\xi y)^3}\delta q(x-\xi y) \right. \nonumber \\
&-& \left. \int_{\frac{x}{\xi}}^{\frac{1+x}{1+\xi}}dy\ \frac{3}{4}\frac{(1+x-\xi y)^2-y^2}{(1+x-\xi y)^3}\delta \bar{q}(-x+\xi y) \right] \nonumber \\
&-& \Theta(-\xi>x)\int_{-\frac{1+x}{1-\xi}}^{\frac{1+x}{1+\xi}}dy\ \frac{3}{4}\frac{(1+x-\xi y)^2-y^2}{(1+x-\xi y)^3}\delta \bar{q}(-x+\xi y)\,.
\end{eqnarray}
For the transversity  PDFs $\delta q$ and $\delta \bar{q}$, we use the parametrization proposed by Anselmino \textit{et al.}~\cite{Anselmino} 

\begin{eqnarray}
\delta u(x) &=& 7.5 \cdot 0.5\cdot (1-x)^5\cdot(x\,u(x)+x\,\Delta u(x))  \,,\\
\delta \bar{u}(x) &=& 7.5 \cdot 0.5\cdot (1-x)^5\cdot(x \,\bar{u}(x)+x\,\Delta \bar{u}(x))  \,,\\
\delta d(x) &=& 7.5\cdot (-0.6)\cdot(1-x)^5\cdot(x\,d(x)+x\,\Delta d(x))  \,,\\
\delta \bar{d}(x) &=& 7.5 \cdot (-0.6) \cdot(1-x)^5\cdot(x \,\bar{d}(x)+x\,\Delta \bar{d}(x)) \,,
\end{eqnarray}
where the helicity-dependent PDFs $\Delta q(x)$, $\Delta\bar{q}(x)$ are parametrized with the help of the unpolarized PDFs $q(x)$ and $\bar{q}(x)$ by \cite{GRSV}

\begin{eqnarray}
\Delta u(x) &=& \sqrt{x}\cdot u(x) \,,\\
\Delta \bar{u}(x) &=& -0.3 \cdot x^{0.4} \cdot\bar{u}(x)  \,,\\
\Delta d(x) &=& -0.7 \cdot\sqrt{x} \cdot d(x)  \,,\\ 
\Delta \bar{d}(x) &=& -0.3\cdot x^{0.4}\cdot\bar{d}(x) \,,
\end{eqnarray}
and the PDFs $q(x)$, $\bar{q}(x)$ come from GRV parametrizations~\cite{GRV}. All these PDFs are calculated at the energy scale $\mu^2 = 10 $ GeV$^2$.
Fig.~\ref{HTu_d_JLab} represents  $H_T^u(x,\xi,t=0)$ and $H_T^d(x,\xi,t=0),$ respectively, for different values of $\xi$, which are determined through (\ref{skewness}) for $S_{\gamma N}=20$ GeV$^2$ of JLab  and for $M^2_{\pi \rho}$ equal 2, 4, 6 GeV$^2$. Similarly, Fig.~\ref{HTu_d_Compass} represents  $H_T^u(x,\xi,t=0)$ and $H_T^d(x,\xi,t=0),$ respectively, for different values of $\xi$, which are determined through (\ref{skewness}) for $S_{\gamma N}=200$ GeV$^2$ of Compass  and for $M^2_{\pi \rho}$ equal 2, 4, 6 GeV$^2$.\\
These two GPDs show some common features like a peak when $x$ is near $\pm \xi$, their order of magnitude and the fact that they both tend to zero when $x$ tends to $\pm 1$. The main difference is their opposite sign. 
The restricted analysis of Ref.~\cite{eps} based on a meson exchange is insufficient for this study since it only gives us the transversity GPDs in the ERBL region. The MIT bag model inspired method of Ref.~\cite{Sco} underestimates the value of  $H_T(x,\xi)$ in the ERBL domain because this model does not take into account antiquark degrees of freedom.  One can notice that these GPDs have the same order of magnitude but some
differences with other models like light-front constituent quark models
\cite{Pasq}, principally due to the fact that in \cite{Pasq}, parametrizations have been
done at $\mu^2$ $\sim$ 0.1 GeV$^2$ whereas our model is calculated at $\mu^2$
$\sim$ 10 GeV$^2$.
 Other model-studies have been developed in the chiral quark soliton model and a QED-based overlap representation \cite{othermodels}.\\

The t-dependence of these chiral-odd GPDs - and its Fourier transform in terms of the transverse
localization of quarks in the proton \cite{impact} - is very interesting but completely unknown. We
will describe it in a simplistic way as:
\begin{equation}
\label{t-dep}
H^q_T(x,\xi,t) = H^q_T(x,\xi,t=0)\times F_H(t),
\end{equation}
where 
\begin{equation}
\label{dipoleFF}
F_H(t) = \frac{C^2}{(t  - C)^2}
\end{equation}
is a standard dipole form factor with  $C=.71~$GeV$^2$. Let us stress that we have no phenomenological control of this assumption, since  the tensor form factor of the nucleon (in fact even the tensor charge) has never been measured.

\newpage

\newpage

\section{Unpolarized Differential Cross Section and Rate Estimates}
\label{Sec:Cross Section and Rates}

\psfrag{ds}{\raisebox{.2cm}{{\hspace{-.3cm}$\left.\frac{d \sigma}{dt \, du' \, d M_{\pi \rho}^2}\right|_{t=t_{min}}$ \hspace{-.3cm}{\footnotesize (nb.GeV$^{-6}$)}}}}
\psfrag{mup}{\raisebox{-.7cm}{{\hspace{-2.5cm}$-u'$(GeV$^2$)}}}
\begin{figure}[h!]
$\begin{array}{cc}
\\
\hspace{-.2cm}
\includegraphics[width=8cm]{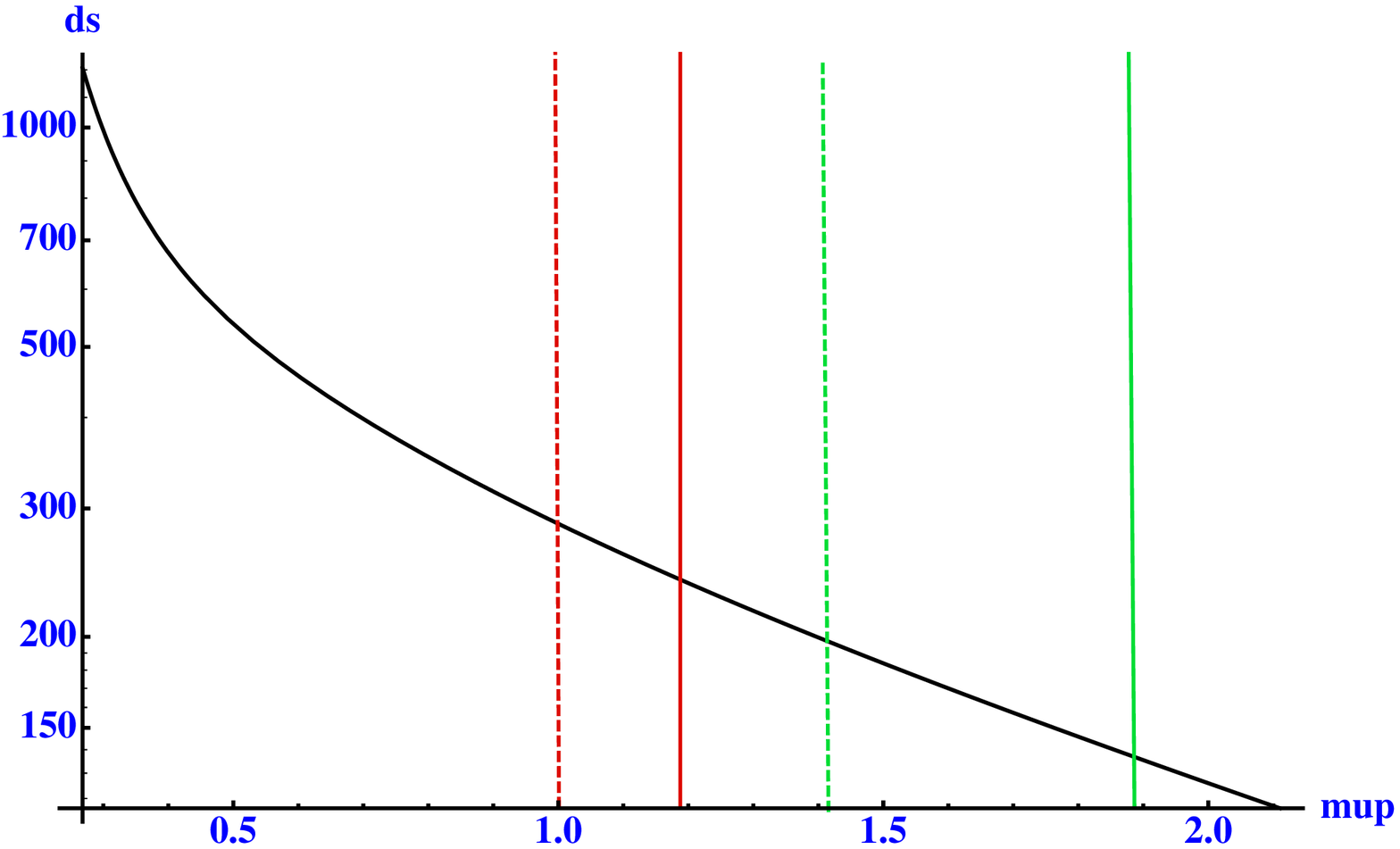}
&
\hspace{-.3cm}
\includegraphics[width=8cm]{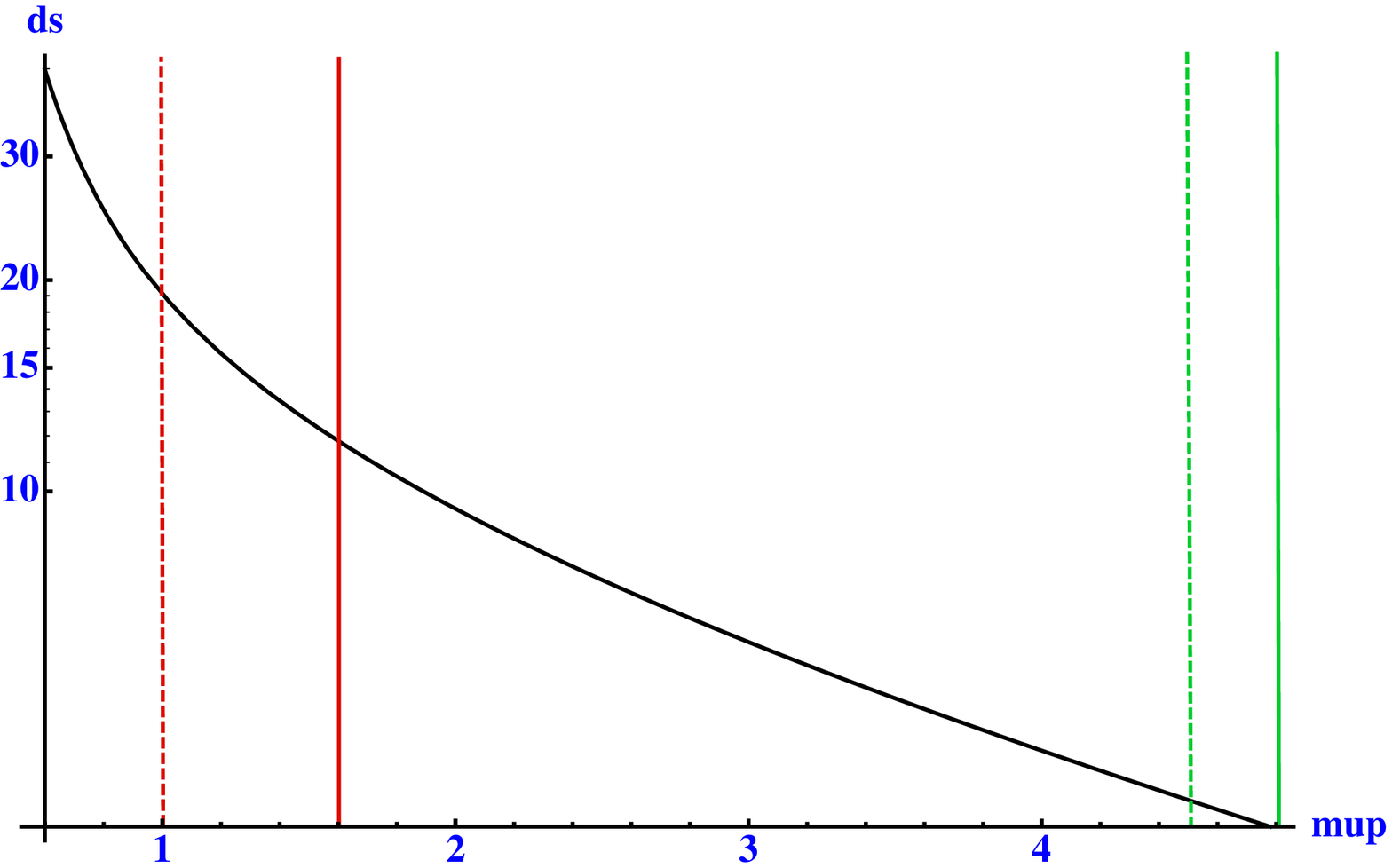}
\\
\\
  (a)    &  (b) \\
\end{array}$
\caption{Variation of the differential cross section (\ref{difcrosec}) (nb.GeV$^{-6}$) with respect to $|u'|$ at $M^2_{\pi\rho}$ = 3 GeV$^2$ (a) and $M^2_{\pi\rho}$ = 6 GeV$^2$ (b) with $S_{\gamma N}$ = 20 GeV$^2$. The lines on the left correspond to the constraints $-u' > 1$ GeV$^2$ and $M_{\pi N'}^2 > 2$ GeV$^2$ and the lines on the right correspond to the constraints $-t' > 1$ GeV$^2$ and $M_{\rho N'}^2 > 2$ GeV$^2$ (dashed line for $t = t_{min}$ and solid line for $t = -0.5$ GeV$^2$).}
\label{resultS20}
\end{figure}

Starting with the expression of the scattering amplitude (\ref{ampl})
 we now calculate the amplitude squared for the unpolarized process
\begin{equation}
\label{amplsqu}
|\mathcal{M}|^2 = \left(\frac{1}{2}\right) \left(\frac{1}{2}\right)\sum_{\lambda_1 \lambda_2} \mathcal{A}\mathcal{A}^* \,.
\end{equation}
It can seem odd to study the chiral-odd quark content of the nucleon by calculating the cross section of an unpolarized scattering but it is enough for now in order to reach this unknown structure. Of course it is possible to consider the polarized one by producing the spin density matrix, which will be done in a future work.

\begin{figure}[h!]
$\begin{array}{cc}
\hspace{-0cm}
\includegraphics[width=8cm]
{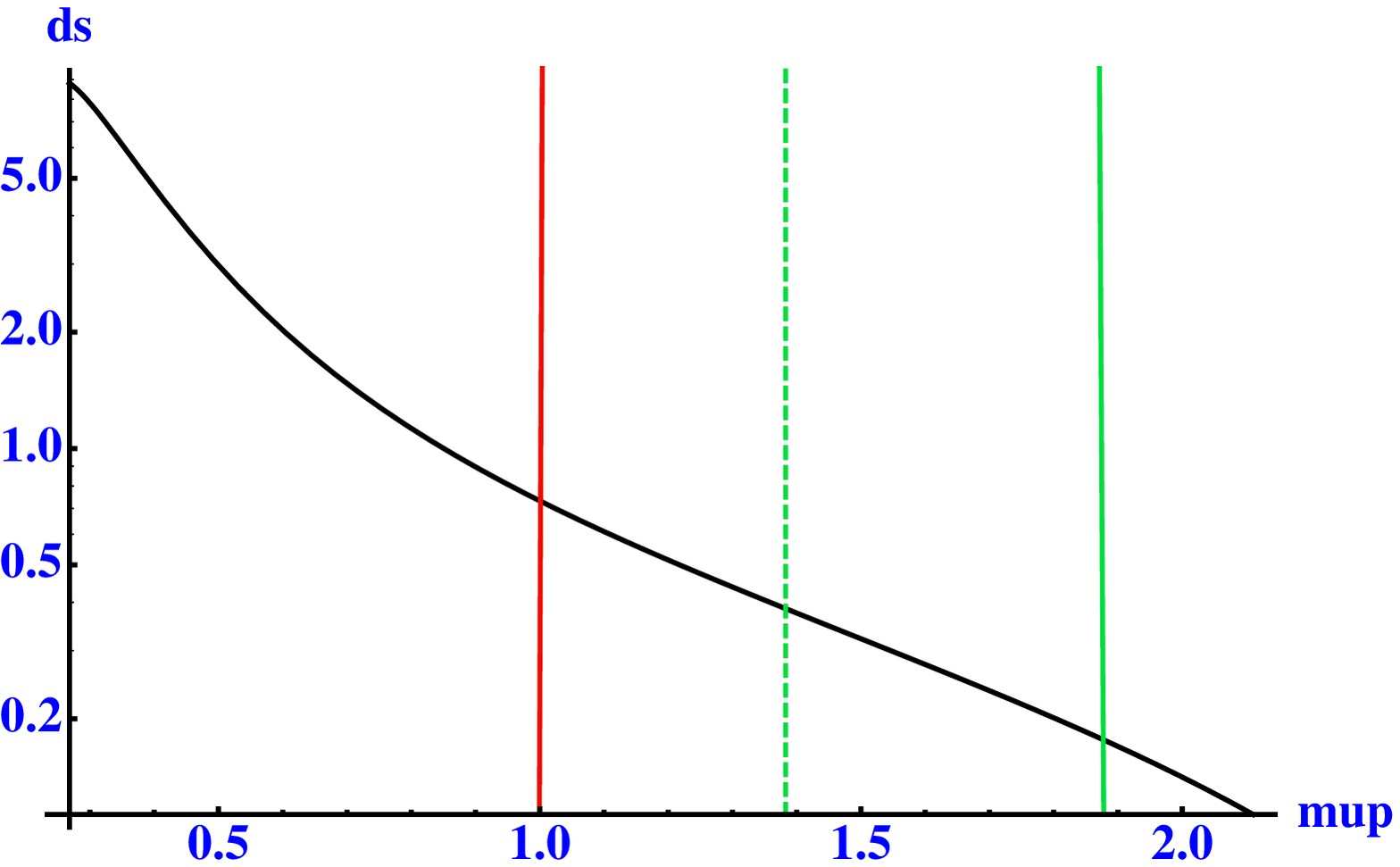}
&
\hspace{-.45cm}
\includegraphics[width=8cm]{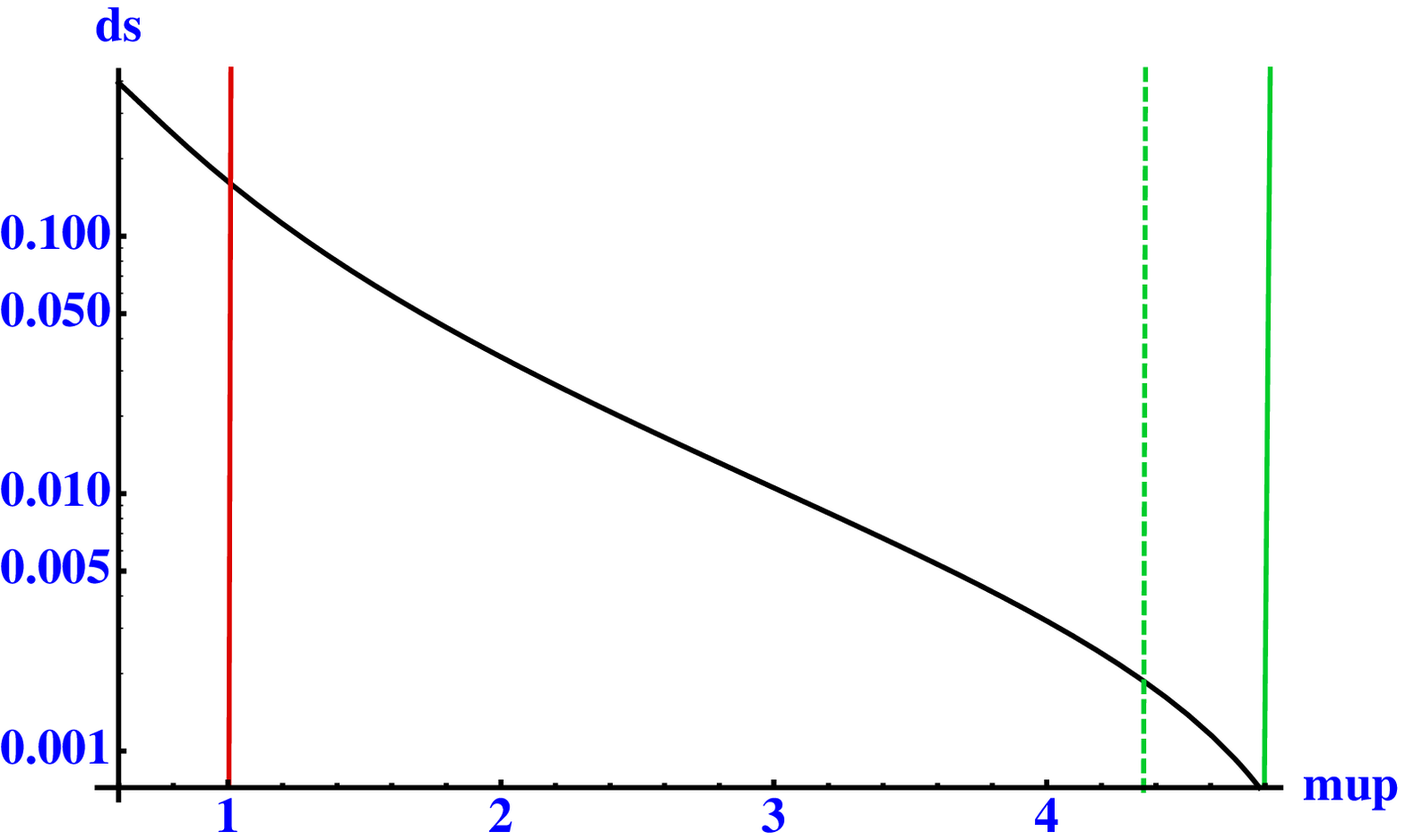}
\\
\\
  (a)    &  (b) \\
\end{array}$
\caption{Variation of the differential cross section (\ref{difcrosec}) (nb.GeV$^{-6}$) with respect to $|u'|$ at $M^2_{\pi\rho}$ = 3 GeV$^2$ (a) and $M^2_{\pi\rho}$ = 6 GeV$^2$ (b) with $S_{\gamma N}$ = 200 GeV$^2$. The solid line on the left corresponds to the constraints $-u' > 1$ GeV$^2$ and $M_{\pi N'}^2 > 2$ GeV$^2$ for any value of $t$ and the lines on the right correspond to the constraints $-t' > 1$ GeV$^2$ and $M_{\rho N'}^2 > 2$ GeV$^2$ (dashed line for $t = t_{min}$ and solid line for $t = -0.5$ GeV$^2$).}
\label{resultS200}
\end{figure}

We now present the cross-section as a function of $t$, $M^2_{\pi\rho},$ $-u'$ which reads
\begin{equation}
\label{difcrosec}
\left.\frac{d\sigma}{dt \,du' \, dM^2_{\pi\rho}}\right|_{\ t=t_{min}} = \frac{|\mathcal{M}|^2}{32S_{\gamma N}^2M^2_{\pi\rho}(2\pi)^3}.
\end{equation}
\noindent
We show, in Fig.~\ref{resultS20},  the differential cross section (\ref{difcrosec}) as a function of  $-u'$ at $S_{\gamma N}$ = 20 GeV$^2$  for $M^2_{\pi\rho}$ = 3 GeV$^2$ i.e. $\xi = 0.085$ and for $M^2_{\pi\rho}$ = 6 GeV$^2$ i.e. $\xi = 0.186$  and, in Fig.~\ref{resultS200}, at $S_{\gamma N}$ = 200 GeV$^2$ respectively for $M^2_{\pi\rho}$ = 3 GeV$^2$ i.e. $\xi = 0.0076$ and for $M^2_{\pi\rho}$ = 6 GeV$^2$ i.e. $\xi = 0.015$.






To get an estimate of the total rate of events of interest for our analysis, we first get the $M^2_{\pi\rho}$ dependence of the differential cross section integrated over $u'$ and $t$,
\begin{equation}
\label{difcrosec2}
\frac{d\sigma}{dM^2_{\pi\rho}} = \int_{(-t)_{min}}^{0.5} \ d(-t)\ \int_{(-u')_{min}}^{(-u')_{max}} \ d(-u') \ F^2_H(t)\times\left.\frac{d\sigma}{dt \, du' d M^2_{\pi\rho}}\right|_{\ t=t_{min}} \,.
\end{equation}

\psfrag{mup}{\hspace{-.2cm}$-u'$}
\psfrag{mt}{$-t$}
\psfrag{mtmin}{$(-t)_{min}$}
\psfrag{mupmax}{\rotatebox{10}{{\hspace{0cm}$(-u')_{max}(t)$}}}
\psfrag{mupmin}{\rotatebox{17}{{\hspace{-1cm}$(-u')_{min(res.)}(t)$}}}
\begin{figure}[!h]
\centerline{\includegraphics[width=8cm]{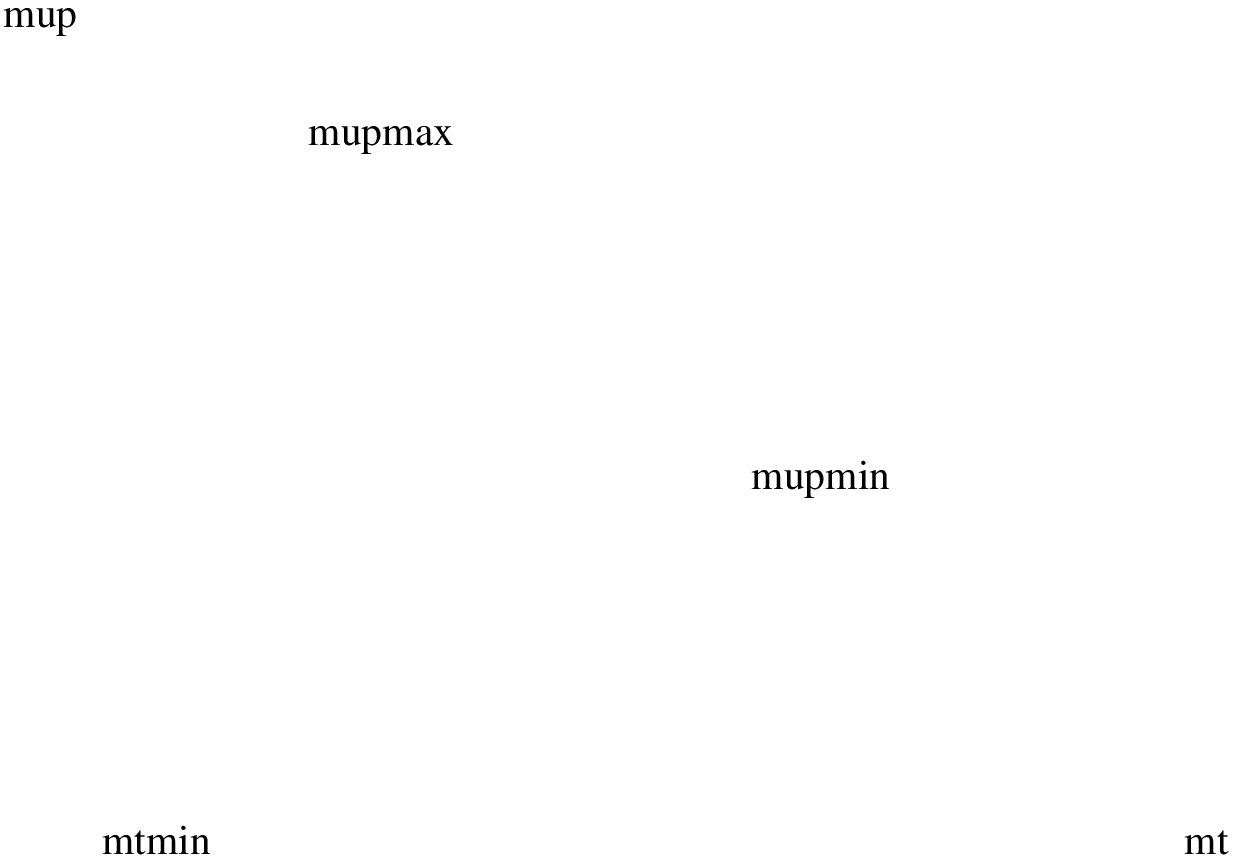}}
\caption{The phase space domain of integration in the $(-t,-u')$ variables. The upper limit in $-u'$ is given by the constraint $-u'(t)<(-u')_{max}(t)=-(-t')_{min}+M_{\pi \rho}^2 -t -m_\pi^2 -m_\rho^2$ (with $(-t')_{min}= 1$ GeV$^2$). The lower limit in  $-u'$ is given
by $-u' > 1$ GeV$^2$ and $-u'(t)>(-u')_{min(res.)}(t)$ where $(-u')_{min(res.)}(t)$ is obtained from the constraint $M_{\pi N'}^2 >$ 2 GeV$^2$. The figure illustrates the case $S_{\gamma N}=20$  GeV$^2$ and $M_{\pi \rho}^2 = 3$ GeV$^2$.}
\label{phase_space}
\end{figure}

The domain of integration over $-u'$ is deduced from the cuts we discussed at the end of section \ref{Sec:Kinematics}: $-t', -u' > 1$ GeV$^2$ and $M_{\pi N'}^2,\ M_{\rho N'}^2 > 2$ GeV$^2$, and is illustrated in Fig.~\ref{phase_space}. Then we get two limits for the domain over $-u'$ :

\begin{itemize}
  \item The cuts over $-u'$ and $M_{\pi N'}^2$ give the minimum value for $-u'$ ($-u'_{min}$ = 1 GeV$^2$ or $(-u')_{min(res.)}$) which depends on $t$, $S_{\gamma N}$ and $M^2_{\pi\rho}$. For instance, the lines on the left in Figs.~\ref{resultS20}-\ref{resultS200} represent that cut : solid lines for $t$ = -0.5 GeV$^2$ and dashed lines for $t = t_{min}$. One can notice that at high energy, the $\pi N'$ system is outside the baryonic resonance region so that $-u'_{min}$ is always equal to 1 GeV$^2$ and for any value of $t$.
  \item The cuts over $-t'$ and $M_{\rho N'}^2$ give the maximum value for $-u'$ ($(-u')_{max}$) which depends on $t$, $S_{\gamma N}$ and $M^2_{\pi\rho}$. For instance, the lines on the right in Figs.~\ref{resultS20}-\ref{resultS200} represent that cut : solid lines for $t$ = -0.5 GeV$^2$ and dashed lines for $t = t_{min}$. It is interesting to stress that, for any value of the hard scale and of the energy, the $\rho N'$ system is always outside the baryonic resonance region.
\end{itemize}

Moreover, one notices that $(-u')_{min}$ decreases and $(-u')_{max}$ increases with $M^2_{\pi\rho}$ at $S_{\gamma N}$ fixed and then the width of the physical region $[(-u')_{min},(-u')_{max}]$ grows.\\

\begin{figure}[!h]
\begin{center}
\includegraphics[width=12cm]{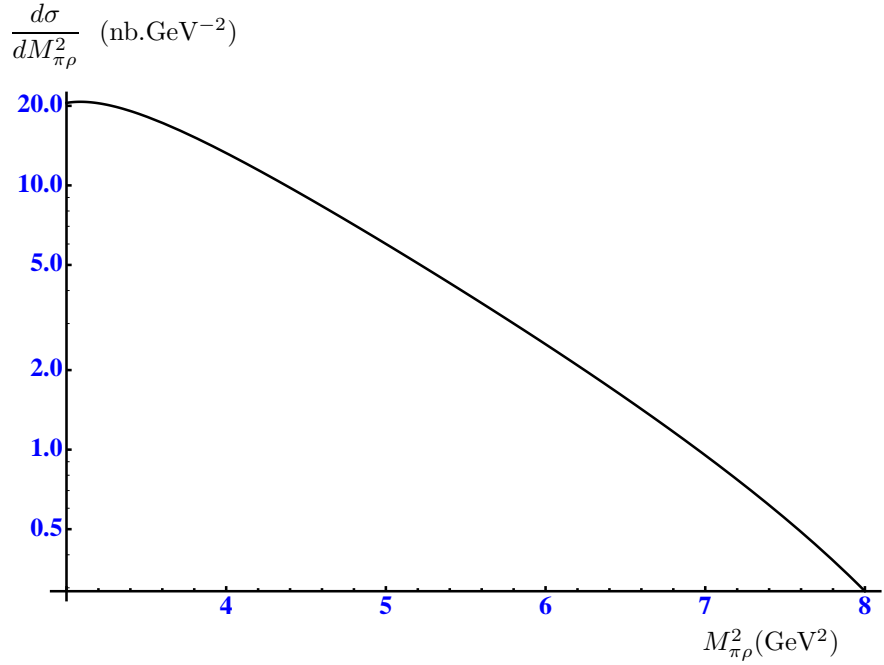}
\vspace{.3cm}
\caption{$M^2_{\pi\rho}$ dependence of the differential cross section (\ref{difcrosec2}) (nb.GeV$^{-2}$) at $S_{\gamma N}$ = 20 GeV$^2$.}
\label{result20}
\end{center}
\end{figure}

\begin{figure}[!h]
\begin{center}
\vspace{.5cm}
\includegraphics[width=12cm]{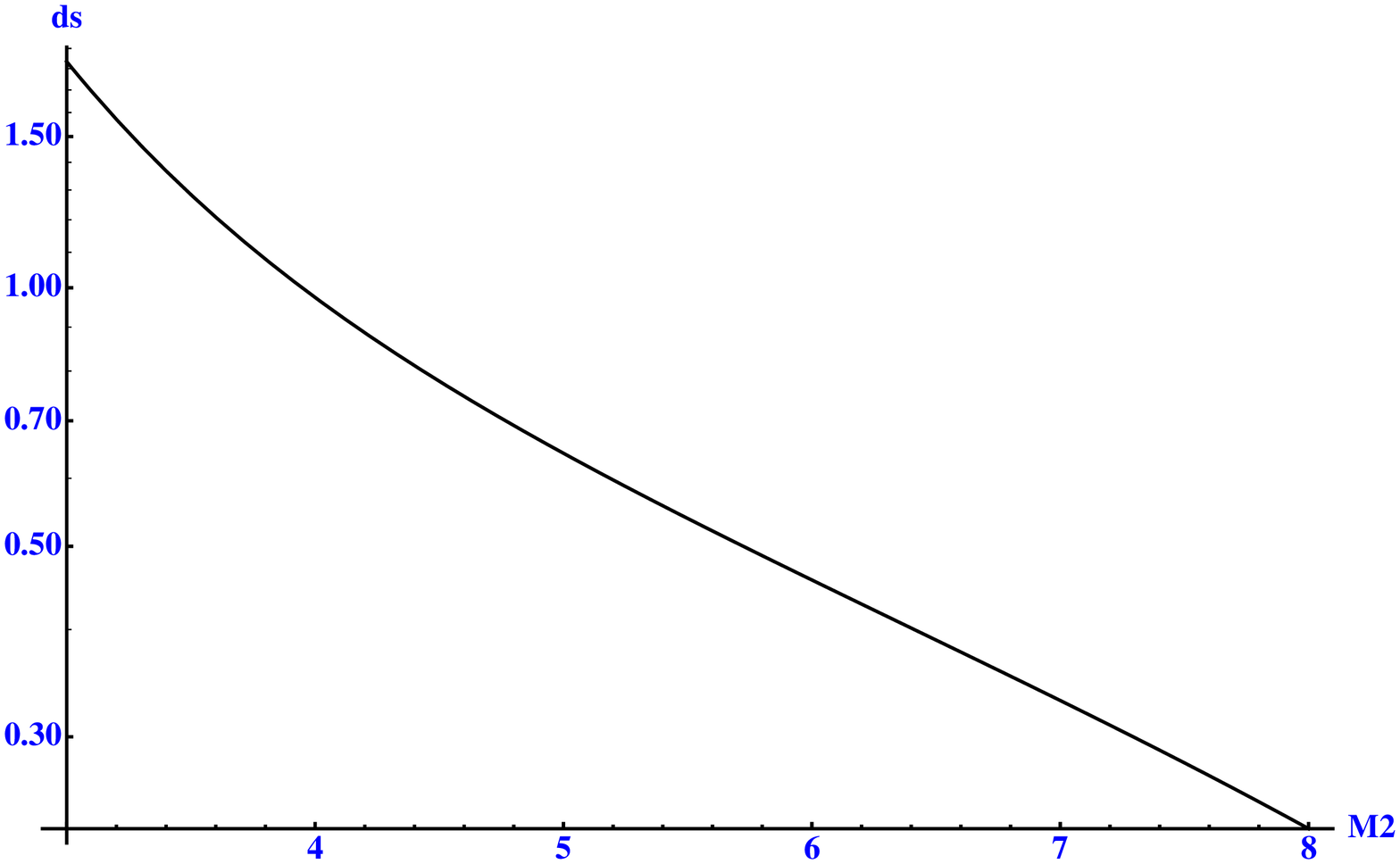}
\vspace{.3cm}
\caption{$M^2_{\pi\rho}$ dependence of the differential cross section (\ref{difcrosec2}) (nb.GeV$^{-2}$) at $S_{\gamma N}$ = 100 GeV$^2$.}
\label{result100}
\end{center}
\end{figure}

\begin{figure}[!h]
\begin{center}
\vspace{.5cm}
\includegraphics[width=12cm]{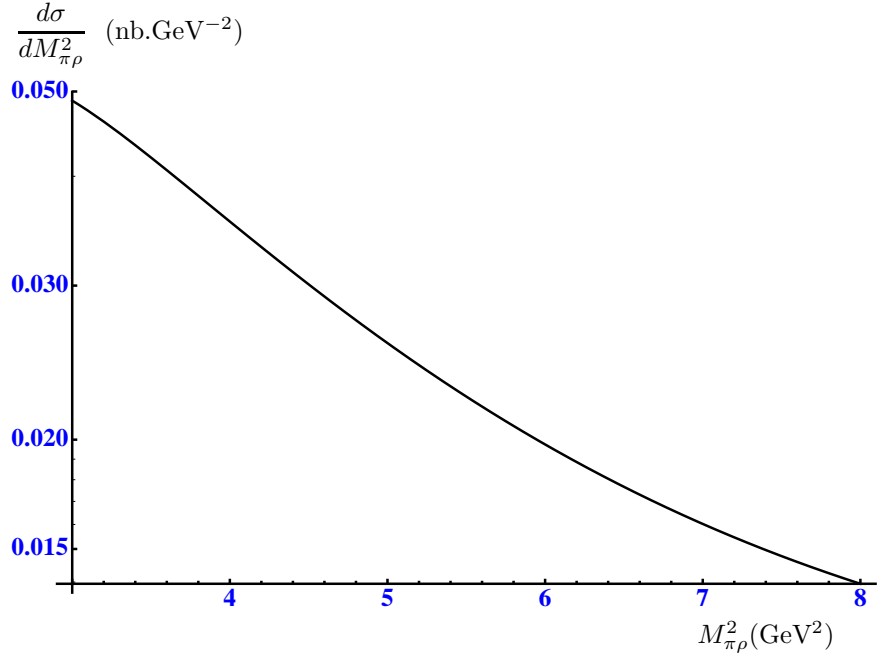}
\vspace{.3cm}
\caption{$M^2_{\pi\rho}$ dependence of the differential cross section (\ref{difcrosec2}) (nb.GeV$^{-2}$) at $S_{\gamma N}$ = 200 GeV$^2$.}
\label{result200}
\end{center}
\end{figure}

Thus, in Figs.~\ref{result20}, \ref{result100} and \ref{result200}, we show the $M^2_{\pi\rho}$ dependence of the differential cross section (\ref{difcrosec2}).

\psfrag{ds}{\raisebox{.5cm}{{\hspace{-.6cm}$\displaystyle \frac{d \sigma}{d M_{\pi \rho}^2}$ \hspace{0cm}{ (nb.GeV$^{-2}$)}}}}
\psfrag{M2}{\raisebox{-.7cm}{{\hspace{-2.5cm}$M_{\pi \rho}^2$(GeV$^2$)}}}


\newpage

Let us first focus on the high energy domain, and discuss the specific case of muoproduction with the COMPASS experiment at CERN. Integrating differential cross sections on $t$, $u'$ and $M^2_{\pi\rho}$, with the cuts specified above  and $M^2_{\pi\rho} > 3$ GeV$^2$,  leads to an estimate of the cross sections for the photoproduction of a $\pi^+\rho^0_T$ pair at high energies such as :
\begin{equation}
\label{crossec1}
\sigma_{\gamma N\to \pi^+\rho^0_TN'}(S_{\gamma N} = 100\ GeV^2) \simeq 3\ \textrm{nb} \qquad \sigma_{\gamma N\to \pi^+\rho^0_TN'}(S_{\gamma N} = 200\ GeV^2) \simeq 0.1\ \textrm{nb}.
\end{equation}
The virtuality $Q^2$ of the exchange photon plays no crucial role in our process, and the virtual photoproduction cross section is almost $Q^2$-independent if we choose to select events in a sufficiently narrow $Q^2-$window (say $Q^2_{min}<Q^2<.5 $ GeV$^2$), which is legitimate since the effective photon flux is strongly peaked at very low values of $Q^2$.  The quasi real (transverse) photon flux $\Gamma_T^l(Q^2,\ \nu) $ reads
\begin{equation}
\label{virphoflux}
\Gamma_T^l(Q^2,\ \nu) = \frac{\alpha \left(\nu - \frac{Q^2}{2M_p}\right)}{2\pi Q^2\nu^2}\left[\left(\frac{\nu}{E_l}\right)^2\left(1 - 2\frac{m^2_l}{Q^2}\right) + \left(1 - \frac{\nu}{E_l} - \frac{Q^2}{4E^2_l}\right)\frac{2}{1 + \frac{Q^2}{\nu^2}}\right]\,,
\end{equation}
with the fine structure constant $\alpha = 1/137$ and $E_l$ the lepton energy (in the laboratory frame).
Consequently, the rate in a photon energy bin $[\nu_1,\nu_2]$ corresponding to  $[\bar S_{\gamma N} - \Delta S,\bar S_{\gamma N} + \Delta S ]$ with $\bar S_{\gamma N} = 2\,\bar \nu \, M = (\nu_1 + \nu_2)M$ and $\Delta S = 2 \, \Delta \nu M = (\nu_2 - \nu_1)M$ is
\begin{eqnarray}
\label{virphoflux2}
\sigma(l \,N\to l\,\pi^+\rho^0_TN') &=& \int_{Q^2_{min}}^{1} dQ^2 \int_{\nu_1}^{\nu_2}d\nu\ \Gamma_T^l(Q^2,\ \nu)\sigma_{\gamma^* N\to \pi^+\rho^0_TN'}(Q^2,\nu)\nonumber\\
&\simeq& \sigma_{\gamma^* N\to \pi^+\rho^0_TN'}(S_{\gamma N} =\bar S_{\gamma N})\times\int_{Q^2_{min}}^{1} dQ^2 \int_{\nu_1}^{\nu_2}d\nu\ \Gamma_T^l(Q^2,\ \nu).
\end{eqnarray}
For the muoproduction ($E_\mu = 160$ GeV), one gets the following cross section estimates, firstly for $S_{\gamma N}$ between 50 and 150 GeV$^2$
\begin{eqnarray}
\label{ }
\sigma(\mu \,N\to \mu \, \pi^+\rho^0_TN') &\simeq& \sigma_{\gamma N\to \pi^+\rho^0_TN'}(S_{\gamma N} = 100\ GeV^2)\times\int_{0.02}^{1} dQ^2 \int_{25}^{75}d\nu\ \Gamma_T^\mu(Q^2,\ \nu)\nonumber\\
&\simeq& 10^{-2}\ \textrm{nb},
\end{eqnarray} 
which yields a rate equal to 3 10$^{-3}$ Hz with a lepton beam luminosity of 2.5 10$^{32}$ cm$^{-2}$.s$^{-1}$, and, for $S_{\gamma N}$ between 150 and 250 GeV$^2$
\begin{eqnarray}
\label{ }
\sigma(\mu \, N\to \mu \, \pi^+\rho^0_TN') &\simeq& \sigma_{\gamma N\to \pi^+\rho^0_TN'}(S_{\gamma N} = 200\ GeV^2)\times\int_{0.02}^{1} dQ^2 \int_{75}^{125}d\nu\ \Gamma_T^\mu(Q^2,\ \nu)\nonumber\\
&\simeq& 5\ 10^{-4}\ \textrm{nb},
\end{eqnarray} 
which yields a rate equal to 1.3 10$^{-4}$ Hz with the same lepton beam luminosity. This looks sufficient to get an estimate of the transversity GPDs in the region of small $\xi$ of the order 0.01.

Let us now turn  to the lower energy domain, which will be studied in details at JLab.
%
%
With the cuts discussed above and $M^2_{\pi\rho} > 3$ GeV$^2$,   estimates of the cross sections for the photoproduction of a $\pi^+\rho^0_T$ pair at JLab energies are:
\begin{equation}
\label{crossec}
\sigma_{\gamma N\to \pi^+\rho^0_TN'}(S_{\gamma N} = 10\ GeV^2) \simeq 15\ \textrm{nb} \qquad \sigma_{\gamma N\to \pi^+\rho^0_TN'}(S_{\gamma N} = 20\ GeV^2) \simeq 33\ \textrm{nb}.
\end{equation}
In electroproduction ($E_e = 11$ GeV), applying Eqs. (\ref{virphoflux}) and (\ref{virphoflux2}), one gets the total cross section
\begin{equation}
\sigma(e^- N\to e^-\pi^+\rho^0_TN') \simeq 0.1\ \textrm{nb}.
\end{equation}
Tagging the photons is however required if one aims at a detailed understanding of the reaction and at an extraction of the GPD. This is possible at JLab and indeed is well documented for the future 12 GeV energy upgrade in \cite{HallDB}. More specifically, Hall D will be equipped with a crystal radiator, which through the technique of coherent brehmsstrahlung will produce an intense photon beam  of 8 - 9 GeV with an excellent degree of polarization. This leads to the following rate
\begin{eqnarray}
\label{rateJLabHallD}
R^D &=& \sigma_{\gamma N\to \pi^+\rho^0_TN'}(S_{\gamma N} = 17\ GeV^2)\times N^D_\gamma\times N^D_p \nonumber\\
&\simeq& 5\ \mathrm{Hz}
\end{eqnarray}
where $N^D_\gamma \sim\ 10^8$ photons/s is the photon flux for Hall D and $N^D_p = 1.27\ b^{-1}$ is the number of protons per surface in the target (liquid hydrogen of 30 cm), assuming that the efficiency of the detector is at 100$\%$.\\ 
With a different technique, CLAS12 in Hall B may be equipped with a photon tagger allowing an intense ($\approx 5\ 10^7 $ photons/s) flux of photons with  energy 7 - 10.5 GeV. This will lead to slightly lower but still large enough rates.\\
Thanks to the high electron beam luminosity expected at JLab, a detailed analysis is possible.\\

Moreover, one can make an additionnal comment about the use of the Compass experiment with kinematics of JLab, i.e. with photons at low energies. In this context, one gets the following cross section estimate for muoproduction for $S_{\gamma N}$ between 20 and 50 GeV$^2$
\begin{eqnarray}
\label{ }
\sigma(\mu N\to \mu\pi^+\rho^0_TN') &\simeq& \sigma_{\gamma N\to \pi^+\rho^0_TN'}(S_{\gamma N} = 35\ GeV^2)\times\int_{0.02}^{1} dQ^2 \int_{10}^{25}d\nu\ \Gamma_T^\mu(Q^2,\ \nu)\nonumber\\
&\simeq& 0.2\ \textrm{nb}\, ,
\end{eqnarray} 
which leads to the conclusion that muoproduction with low energy, at Compass, gives greater rates (5 10$^{-2}$ Hz) than with photon with high energy.

\section{Conclusions}

In this paper, we have advocated  that the exclusive photoproduction of a   meson pair with a large invariant mass gives access to  generalized parton distributions through the factorization of a hard subprocess, provided all the kinematical invariants ($s',t',u'$) which characterize this subprocess are large enough.  We applied this strategy to access the chiral-odd  generalized quark distributions from the photoproduction of a  $\pi^+ \rho^0_T$ meson pair with a large invariant mass. We modeled the dominant chiral-odd GPD $H^{ud}_T(x,\xi,t)$ though a double distribution constrained by the phenomenological knowledge of the transversity quark distribution $h_1^u(x)$ and $h_1^d(x)$. The calculation of the hard part at the leading order in the strong coupling $\alpha_s$ shows that no divergence nor end-point singularity plagues the validity of our approach. From our results, we conclude that the experimental search is promissing, both at low real or almost real photon energies within the JLab@12GeV upgraded facility, with the nominal  effective luminosity generally expected ($\mathcal{L} \sim 10^{35}$ cm$^2$.s$^{-1}$) and at higher photon energies with the  Compass experiment at CERN. These two energy ranges should give complementary information on the chiral-odd GPD $H_T(x,\xi,t)$. Namely, the large $\xi$ region may be scrutinized at JLab and the smaller $\xi$ region may be studied at COMPASS.

It is obvious that our model for the chiral-odd GPD $H_T^q$ may be improved and refined in many ways,  for instance by adding a D-term which gives a complete parametrization by double distribution. Our primary goal in this paper was to prove the feasibility of the  study of this physics with this physical process. We believe that this task is achieved.

The described framework opens the way to future studies. Firstly, the contributions proportionnal to  the other three chiral-odd GPDs should be included in the calculation of the amplitude. Alhough they are suppressed by kinematical factors at small $t$, they constitute an interesting addition to the transversity structure of the nucleon.  Secondly, it will be interesting to study the polarized process and calculate more observables like beam or target spin asymmetries,  or like the spin density matrix of the vector meson. These two extensions will be discussed in a separate paper. Other two meson channels may also be interesting; they deserve a thorough study.

\section*{Acknowledgments}

We are grateful to Igor Anikin, Markus Diehl, Samuel Friot, Franck Sabatie and Jean Philippe Lansberg for useful 
discussions and correspondance. 
This work is partly supported by the French-Polish scientific agreement Polonium 7294/R08/R09,  the ECO-NET program, contract 18853PJ, the ANR-06-JCJC-0084-02, the Polish Grant N202 249235 and the DFG (KI-623/4).


\end{document}